\documentclass[final,3p,number,sort&compress]{elsarticle}
\usepackage{lineno}

\usepackage{soul}
\newif\ifdebug
\debugfalse

\newif\ifanon
\anonfalse

\usepackage[hidelinks]{hyperref}
\usepackage{doi}
\usepackage{amssymb}
\usepackage{amsmath}
\usepackage{amsthm,xparse}
\usepackage{paralist}
\ifanon \journal{Anonymous Journal} \else \journal{Theoretical Computer Science}\fi

\usepackage[american]{babel}
\usepackage[utf8]{inputenc}
\usepackage{csquotes}
\usepackage{listings}
\usepackage{rotating}
\usepackage[table]{xcolor}
\usepackage{tabularx,booktabs}
\usepackage{multirow}
\usepackage{slashbox}
\usepackage[graphicx]{realboxes}

\usepackage{amssymb}
\usepackage{amsmath}
\usepackage{dblfloatfix}
\usepackage{paralist}
\usepackage{subcaption}
\usepackage{bm}
\usepackage{url}

\usepackage[ruled,linesnumbered,noend]{algorithm2e}
\usepackage{subcaption}

\usepackage[colorinlistoftodos,prependcaption,textsize=small]{todonotes} 

\makeatletter
\def\Url@twoslashes{\mathchar`\/\@ifnextchar/{\kern-.2em}{}}
\g@addto@macro\UrlSpecials{\do\/{\Url@twoslashes}}
\makeatother
\makeatletter
\g@addto@macro{\UrlBreaks}{\UrlOrds}
\makeatother
\usepackage[capitalise,nameinlink]{cleveref}
\crefname{section}{Section}{Sections}
\crefname{figure}{Figure}{Figures}
\crefname{example}{Example}{Examples}
\crefname{algorithm}{Algorithm}{Algorithms}

\usepackage{rotating}

\usepackage{lscape}

\newcommand{\rotateHeading}[1]{\rotatebox[origin=b]{90}{\textbf{#1}}}

\newcommand{\images}{figures/}

\newcommand{\eg}{e.\,g.,\ }
\newcommand{\ie}{i.\,e.,\ }
\newcommand{\etal}{et~al.}

\ifanon
\newcommand{\model}{ANONYMOUS formal language}
\newcommand{\modelLong}{Hidden in plain sight}
\else
\newcommand{\model}{FLUID}
\newcommand{\modelLong}{FLexible graph sUmmarIes for Data graphs}
\fi
\newcolumntype{L}[1]{>{\Centering\arraybackslash}p{#1}}
\newcolumntype{Y}[2]{*{#1}{>{\hsize=#2\hsize\centering\arraybackslash}X}}
\mathchardef\mhyphen="2D
\newcommand{\OR}{\mathbin{\cup_{\mathrm{ex}}}}
\mathchardef\<="3C
\mathchardef\>="3E

\newcommand{\id}{\mathrm{id}}
\newcommand{\idrel}{\mathrm{id_{rel}}}
\newcommand{\indir}{\mathrm{i}}
\newcommand{\outdir}{\mathrm{o}}
\newcommand{\bidir}{\mathrm{b}}
\newcommand{\taut}{\top}
\newcommand{\EQR}{\mathrm{EQR}}
\newcommand{\PAY}{\mathrm{PAY}}
\newcommand{\PC}{\mathrm{PC}}
\newcommand{\OC}{\mathrm{OC}}
\newcommand{\POC}{\mathrm{POC}}
\newcommand{\OCtype}{\OC_{\mathrm{type}}}
\newcommand{\PCrel}{\PC_{\mathrm{rel}}}
\newcommand{\ts}[1]{\ell_T(#1)}
\newcommand{\ps}[1]{\ell_P(#1)}
\newcommand{\psin}[1]{\ell_P^-(#1)}
\newcommand{\nsout}[1]{\Gamma^+(#1)}
\newcommand{\nsin}[1]{\Gamma^-(#1)}
\newcommand{\datasourcepay}{dsp}
\newcommand{\countpay}{vcp}
\newcommand{\vertexpay}{vip}
\newcommand{\PRDFS}{P_{\mathrm{RDFS}}}
\newcommand{\VGRDFS}{VG_{\mathrm{RDFS}}}
\newcommand{\simple}{\lambda}
\newcommand{\complex}{\kappa}
\newcommand{\timbl}{TimBL-11M}
\newcommand{\dyldo}{DyLDO-127M}
\newcommand{\btc}{BTC-2B}
\newcommand{\laundromat}{Laundromat-38B}

\newcommand{\lastaccess}{last accessed: October 16, 2020}

\newdefinition{definition}{Definition}

\newtheorem{lemma}{Lemma}
\newdefinition{remark}{Remark}

\newdefinition{example}{Example}

\makeatletter
\newcommand{\myref}[1]{\mynameref{#1}{\csname r@#1\endcsname}\space(\cref{#1})}
\newcommand{\Myref}[1]{\Cref{#1}\mynameref{#1}{\csname r@#1\endcsname}}

\def\mynameref#1#2{%
  \begingroup
    \edef\@mytxt{#2}%
    \edef\@mytst{\expandafter\@thirdoffive\@mytxt}%
    \ifx\@mytst\empty\else
    \space\nameref{#1}\fi
  \endgroup
}
\makeatother

\colorlet{punct}{red!60!black}
\definecolor{background}{HTML}{EEEEEE}
\definecolor{delim}{RGB}{20,105,176}
\colorlet{numb}{magenta!60!black}
\lstdefinelanguage{json}{
    basicstyle=\normalfont\ttfamily,
    numbers=left,
    numberstyle=\scriptsize,
    stepnumber=1,
    numbersep=8pt,
    showstringspaces=false,
    breaklines=true,
    frame=lines,
    backgroundcolor=\color{background},
    literate=
     *{0}{{{\color{numb}0}}}{1}
      {1}{{{\color{numb}1}}}{1}
      {2}{{{\color{numb}2}}}{1}
      {3}{{{\color{numb}3}}}{1}
      {4}{{{\color{numb}4}}}{1}
      {5}{{{\color{numb}5}}}{1}
      {6}{{{\color{numb}6}}}{1}
      {7}{{{\color{numb}7}}}{1}
      {8}{{{\color{numb}8}}}{1}
      {9}{{{\color{numb}9}}}{1}
      {:}{{{\color{punct}{:}}}}{1}
      {,}{{{\color{punct}{,}}}}{1}
      {\{}{{{\color{delim}{\{}}}}{1}
      {\}}{{{\color{delim}{\}}}}}{1}
      {[}{{{\color{delim}{[}}}}{1}
      {]}{{{\color{delim}{]}}}}{1},
}

\lstdefinelanguage{bnf}{
    basicstyle=\normalfont\ttfamily,
    numbers=left,
    numberstyle=\scriptsize,
    stepnumber=1,
    numbersep=8pt,
    showstringspaces=false,
    breaklines=true,
    frame=lines,
    backgroundcolor=\color{background},
    literate=
     *{0}{{{\color{numb}0}}}{1}
      {1}{{{\color{numb}1}}}{1}
      {2}{{{\color{numb}2}}}{1}
      {3}{{{\color{numb}3}}}{1}
      {4}{{{\color{numb}4}}}{1}
      {5}{{{\color{numb}5}}}{1}
      {6}{{{\color{numb}6}}}{1}
      {7}{{{\color{numb}7}}}{1}
      {8}{{{\color{numb}8}}}{1}
      {9}{{{\color{numb}9}}}{1}
      {;}{{{\color{delim}{;}}}}{1}
      {|}{{{\color{delim}{$\mid$}}}}{1}
      {\{}{{{\color{delim}{\{}}}}{1}
      {\}}{{{\color{delim}{\}}}}}{1}
      {[}{{{\color{delim}{[}}}}{1}
      {]}{{{\color{delim}{]}}}}{1},
}

\begin{document}
\lefthyphenmin=4
\righthyphenmin=4
\sloppy
\tolerance=1000

{
\ifdebug
\pagestyle{empty}
\onecolumn
\tableofcontents
\cleardoublepage
\fi
}
\setcounter{page}{1}
\begin{frontmatter}


\title{FLUID: A Common Model for Semantic Structural Graph Summaries\\ Based on Equivalence Relations}

\ifanon \author{Anonymous author(s)} \else
\author[kiel,ey]{Till Blume}
\author[essex]{David Richerby}
\author[essex,ulm]{Ansgar Scherp}

\address[kiel]{Kiel University, Germany}
\address[ey]{Ernst \& Young GmbH WPG – GSA R\&D, Germany}

\address[essex]{University of Essex, UK}
\address[ulm]{Ulm University, Germany}
\fi

\begin{abstract}
Summarization is a widespread method for handling very large graphs.
The task of structural graph summarization is to compute a concise but meaningful synopsis of the key structural information of a graph.
As summaries may be used for many different purposes, there is no single concept or model of graph summaries.
We have studied existing structural graph summaries for large-scale (semantic) graphs.
Despite their different concepts and purposes, we found commonalities in the graph structures they capture.
We use these commonalities to provide for the first time a formally defined common model, \model{} (\modelLong{}), that allows us to flexibly define structural graph summaries.
\model{} allows graph summaries to be quickly defined, adapted, and compared for different purposes and datasets.
To this end, \model{} provides features of structural summarization based on equivalence relations such as distinction of types and properties, direction of edges, bisimulation, and inference.
We conduct a detailed complexity analysis of the features provided by \model{}.
We show that graph summaries defined with \model{} can be computed in the worst case in time $\mathcal{O}(n^2)$ w.r.t.\@ $n$, the number of edges in the data graph.
An empirical analysis of large-scale web graphs with billions of edges indicates a typical running time of $\Theta(n)$.
Based on the formal \model{} model, one can quickly define and modify various structural graph summaries from the literature and beyond.
\end{abstract}


\end{frontmatter}


\section{Introduction}
\label{sec:introduction}
Representing data as a graph is increasingly popular, since graphs allow a more efficient and a more flexible implementation of certain applications compared to traditional relational databases~\cite{DBLP:journals/corr/abs-2003-02320}.
In general, such data graphs contain labeled vertices and edges, which describe the data and their relationships. 
However, in huge, heterogeneous data graphs like the Linked Open Data cloud,\footnote{\url{https://lod-cloud.net/}, \lastaccess.} several tasks are computationally expensive, such as cardinality computations for queries~\cite{DBLP:conf/icde/NeumannM11},
data exploration~\cite{lodex2015,loupe2015,DBLP:conf/semWeb/PietrigaGADCGM18,DBLP:conf/esws/SpahiuPPRM16a}, data visualization~\cite{DBLP:journals/vldb/GoasdoueGM20},
vocabulary term recommendations~\cite{DBLP:conf/esws/SchaibleGS16}, and related entity retrieval~\cite{DBLP:conf/www/CiglanNH12}.
%
Graph summarization facilitates the identification of meaning and structure in data graphs~\cite{DBLP:journals/csur/LiuSDK18}.
Thus, graph summaries can be used to tackle the above tasks more efficiently, \eg by serving as indices.

Different attempts have been made to classify the different graph summarization approaches~\cite{DBLP:journals/csur/LiuSDK18,DBLP:journals/corr/abs-2004-14794,DBLP:journals/pvldb/KhanBB17,DBLP:journals/vldb/CebiricGKKMTZ19}.
%
In this work, we focus on structural graph summaries, \ie graph summaries that precisely capture specific structural features of the data graph~\cite{DBLP:journals/vldb/CebiricGKKMTZ19}.
Structural graph summaries are condensed representations of graphs such that a set of chosen (structural) features of the graph summary are equivalent to the original graph.
To achieve this, structural graph summaries partition vertices based on equivalent subgraphs.
To determine subgraph equivalences, only structural features are used, such as specific combinations of labels.
Structural graph summaries are usually an order of magnitude smaller than the input graph but are equivalent to the original graph regarding the chosen structural features~\cite{DBLP:conf/dexa/BlumeS20}.
Structural graph summaries do not include statistical approaches such as sampling or pattern-mining approaches~\cite{DBLP:journals/vldb/CebiricGKKMTZ19}.

Semantic structural graph summaries extend structural graph summaries by supporting concepts from ontologies, such as semantic labels for vertices and edges~\cite{DBLP:journals/vldb/CebiricGKKMTZ19,DBLP:conf/esws/SpahiuPPRM16a,DBLP:journals/ws/KonrathGSS12}.
Ontologies model hierarchical relationships between concepts, \ie relationships between types and properties with which the vertices and edges are labeled, \eg the type \texttt{Proceedings} is a subtype of the type \texttt{Book}.
Following this subtype relation allows us to infer the type \texttt{Book} for all vertices that are labeled with the type \texttt{Proceedings}.

Many different structural graph summaries have been developed for different purposes, capturing different structural features of graphs~\cite{loupe2015,lodex2015,DBLP:conf/dexaw/CampinasPCDT12,DBLP:conf/icde/NeumannM11,DBLP:conf/www/CiglanNH12,DBLP:journals/ws/KonrathGSS12,DBLP:conf/esws/SpahiuPPRM16a,DBLP:conf/esws/SchaibleGS16,DBLP:journals/vldb/GoasdoueGM20,DBLP:journals/tkde/TranLR13,DBLP:conf/icde/KaushikSBG02,DBLP:conf/icdt/MiloS99,DBLP:journals/pvldb/ConsensFKP15,DBLP:conf/sigmod/SchatzleNLP13}.
The problem with this plethora of existing summary models is that each model defines its own data structure that is designed for solving only a specific task~\cite{DBLP:conf/icde/NeumannM11,lodex2015,loupe2015,DBLP:conf/semWeb/PietrigaGADCGM18,DBLP:conf/esws/SpahiuPPRM16a,DBLP:conf/esws/SchaibleGS16,DBLP:conf/www/CiglanNH12,DBLP:conf/kcap/GottronSKP13}.
Our observation is that the different tasks cannot be sufficiently supported by any of the existing structural graph summaries~\cite{DBLP:conf/dexa/BlumeS20}.
Furthermore, it is difficult to compare structural graph summaries, as they were designed, implemented, and evaluated in isolation, for their individual task only and using different queries, datasets, graph models, and metrics.

However, when analyzing the fundamental concepts underlying the various existing (semantic) structural graph summaries, one can identify a common theoretical grounding.
Either explicitly, or often only implicitly, the existing graph summary models are defined using equivalence relations~\cite{DBLP:conf/dexaw/CampinasPCDT12,DBLP:conf/icde/NeumannM11,DBLP:conf/www/CiglanNH12,DBLP:journals/ws/KonrathGSS12,DBLP:conf/esws/SpahiuPPRM16a,lodex2015,loupe2015,DBLP:conf/esws/SchaibleGS16,DBLP:journals/vldb/GoasdoueGM20,DBLP:journals/tkde/TranLR13,DBLP:conf/icde/KaushikSBG02,DBLP:conf/icdt/MiloS99,DBLP:journals/pvldb/ConsensFKP15,DBLP:conf/sigmod/SchatzleNLP13}.
%
This paper summarizes our extensive, in-depth analysis of the existing (semantic) structural graph summaries.
From this analysis, we have identified common features in existing structural graph summaries.
We model these features as equivalence relations, which can be flexibly combined to form new equivalence relations, and by this be tailored for each individual purpose.
The result is a small set of operators that make up our common model \model{} (short for \modelLong{}), which defines a language for flexibly defining semantic structural graph summaries.
%
%
Since only a few operators are needed to define these equivalence relations, we are able to produce a single, parameterized algorithm that computes all structural graph summaries defined with \model{}.
Any graph summary defined with \model{} can be computed in worst case in time $\mathcal{O}(n^2)$ w.r.t.\@ $n$, the number of edges in the data graph.
Analyzing large-scale graphs on the web with billions of edges shows that, empirically, their computation time is in $\Theta(n)$.


%

In summary, the contributions of this work are:
\begin{enumerate}[(I)]
\item A formally defined language \model{} for flexibly defining and adapting structural graph summaries and semantic structural graph summaries.

\item A parameterized, sequential algorithm that computes all graph summaries defined with \model{}.

\item The worst case time complexity of computing summaries with \model{} is in $\mathcal{O}(n^2)$ w.r.t.\@ $n$, the number of edges in the input graph.
An empirical analysis of large-scale web graphs with billions of edges indicates a typical computation time of $\Theta(n)$.
\end{enumerate}
By introducing \model{}, we lay the foundation for a new generation of flexibly defined and easy to use semantic structural graph summaries.
Our approach scales for complex tasks and large-scale semantic graphs with billions of edges.
Furthermore, the implementation of our graph summarization framework is freely available.\footnote{\url{https://github.com/t-blume/fluid-framework}}

This paper is an extension of our previous workshop paper~\cite{DBLP:conf/gvd/2018}. 
The previous work defines \model's schema elements and five parameterizations and conducts a first complexity analysis. 
In this work, we extend the functionality of \model{} by introducing the new set parameterization and the related property instance parameter for the instance parameterization.
Additionally, we formally introduce the notion of payload to \model{}, which is needed to adapt graph summaries to different application tasks.
Furthermore, we provide an extended literature review including a detailed description of the features provided by related works.
We also provide explanations and examples for all schema elements and their parameterizations.
We also present our single, parameterized algorithm and provide an extensive and more precise complexity analysis of our algorithm.
Finally, we also included experiments to estimate the impact of the inference parameterization in typical scenarios. 

The remainder of this paper is structured as follows. In the next section, we describe our problem statement in detail.
We analyze existing (semantic) structural graph summaries in \cref{sec:related-work}.
In \cref{sec:formal-defintions}, we formally define the \model{} language and show how to define structural graph summaries.
We present our parameterized algorithm to compute structural graph summaries defined with \model{} and analyze its complexity in \cref{sec:complexity}.
Finally, in \cref{sec:dataset-analysis}, we analyze four large-scale, real-world datasets with billions of edges to substantiate that the typical computational complexity is $\Theta(n)$, before we conclude in \cref{sec:conclusion}.

\section{Problem Statement}
\label{sec:problem-statement}
Structural graph summarization is the task of finding a condensed representation $SG$ (short for ``summary graph'') of an input graph $G$ such that a set of chosen (structural) features are equivalent in $SG$ and the original graph~\cite{DBLP:journals/vldb/CebiricGKKMTZ19}.
Intuitively, structural graph summarization means that we can conduct specific tasks~-- \eg counting the vertices with a specific type label~-- directly on $G$ or, alternatively, obtain the same information from the structural graph summary $SG$.
The fundamental idea of structural graph summaries is that the task can be completed much faster on the graph summary than on the original graph.
In this paper, without loss of generality, we consider Resource Description Framework (RDF) graphs as input graphs.
We chose RDF graphs due to their popularity and because they are standardized by the World Wide Web Consortium (W3C)~\cite{w3c-rdf}.\footnote{The Linked Open Data cloud, available at \url{https://lod-cloud.net/}, is a popular depiction of the use of RDF graphs on the web since May 2007. The current depiction of the cloud from July 2020 formally registers 1,260 datasets stemming from different organizations and domains.}
Below, we formally introduce the RDF graph model.
Subsequently, we introduce the notion of semantic structural graph summaries.

\subsection{RDF Graphs}
\label{sec:graph-model}
Conceptually, an RDF graph is an edge-labelled directed graph. 
An RDF graph~\cite{w3c-rdf} is a set of triples $(s, p, o)$, with a subject~$s$, predicate~$p$, and object~$o$.
Each triple denotes a directed edge from the subject vertex~$s$ to the object vertex~$o$, the edge being labeled with the predicate~$p$.
RDF graphs distinguish three kinds of vertices, namely International Resource Identifiers (IRIs)~\cite{rfc-iri}, blank nodes, and literals, each of which has a different role.

Formally, an RDF graph is defined as $G \subseteq (V_I\cup V_{B}) \times P \times (V_I \cup V_{B} \cup L)$, where $V_I$ denotes the set of IRIs, $V_{B}$ the set of blank nodes, $P \subseteq V_I$ the set of predicates (also identified by IRIs), and $L$ the set of literals (represented by strings)~\cite{w3c-rdf}.
IRIs conceptually correspond to real-world entities and are globally unique: an IRI may be included in more than one RDF graph, but this corresponds to stating different facts about the same real-world entity. 
In contrast, blank nodes are only locally defined, within the scope of a specific RDF graph, to serve special data modeling tasks.
Through skolemization, blank nodes can be turned into Skolem IRIs, which are globally unique~\cite[Section~3.5]{w3c-rdf}. 
Literal vertices are finite strings of characters from a finite alphabet such as Unicode~\cite{w3c-rdf}.
Thus, two literal vertices which are term-equal (\ie are the same string)~\cite[Section~3.3]{w3c-rdf} are the same vertex.
Although IRIs, blank nodes, and literals have different roles in RDF, this distinction is not relevant in this work and we treat them equally.

Predicates $p \in P$ act as edge labels. 
The RDF standard includes the predicate \texttt{rdf:type}, which is used to simulate vertex labels: the triple $(s, \texttt{rdf:type}, o)$ denotes the vertex~$s$ having label~$o$. 
Such vertex labels are called RDF types.
This indirect representation of vertex labels is a design decision of RDF and contrasts with the direct use of vertex labels in labeled property graphs~\cite{DBLP:journals/corr/abs-2003-02320}.
Edges in an RDF graph which are not labeled with \texttt{rdf:type} are called RDF properties.

We define the set $\ts{s} := \{o \in V_I \mid (s, \texttt{rdf:type}, o)\in G\}$ as the \textbf{type set} of a vertex~$s$ in an RDF graph~\cite{DBLP:journals/dpd/GottronKS15}.
We define the set $\ps{s} := \{p \in P \mid (s,p,o)\in G\text{ for some }o\text{ and }p\neq \texttt{rdf:type}\}$ as the \textbf{property set} of a vertex $s$ in an RDF graph~\cite{DBLP:journals/dpd/GottronKS15}.
We write $V_C$ for the set of all RDF types, \ie the set $\{o\in V_I\mid (s, \texttt{rdf:type}, o) \in G \text{ for some }s\}$.
Furthermore, we write $\nsout{s} = \{o \mid (s,p,o)\in G\text{ for some }p\}$ for the set of outgoing neighbors of~$s$ in the RDF graph~$G$ and $\nsin{o} = \{s \mid (s,p,o)\in G\text{ for some }p\}$ for the set of $o$'s incoming neighbors.
Finally, we define $\psin{o} := \{p \mid (s,p,o) \in G \text{ for some } s \text{ and } p \not= \texttt{rdf:type}\}$ as the \textbf{incoming property set}.

\begin{figure}[h]
    \centering
\begin{minipage}[t]{.48\textwidth}
    \centering
    \vspace*{0.5cm}
   \begin{align*}  
  \{&(v_1,\texttt{rdf:type},\texttt{Proceedings}),\\
    &(v_2,\texttt{rdf:type},\texttt{Person}),\\
    &(v_1,\texttt{author},v_2),\\
    &(v_1,\texttt{title},\text{\enquote{Graph Database}}),\\
    &(v_2,\texttt{name},\text{\enquote{Max Power}}) \}
	\end{align*}  
	\vspace*{0.5cm}
\end{minipage}%
\quad
\begin{minipage}[b]{0.48\textwidth}
    \centering 
    \includegraphics[scale=.6,trim={33cm 16.5cm 27.5cm 10cm}, clip=true]{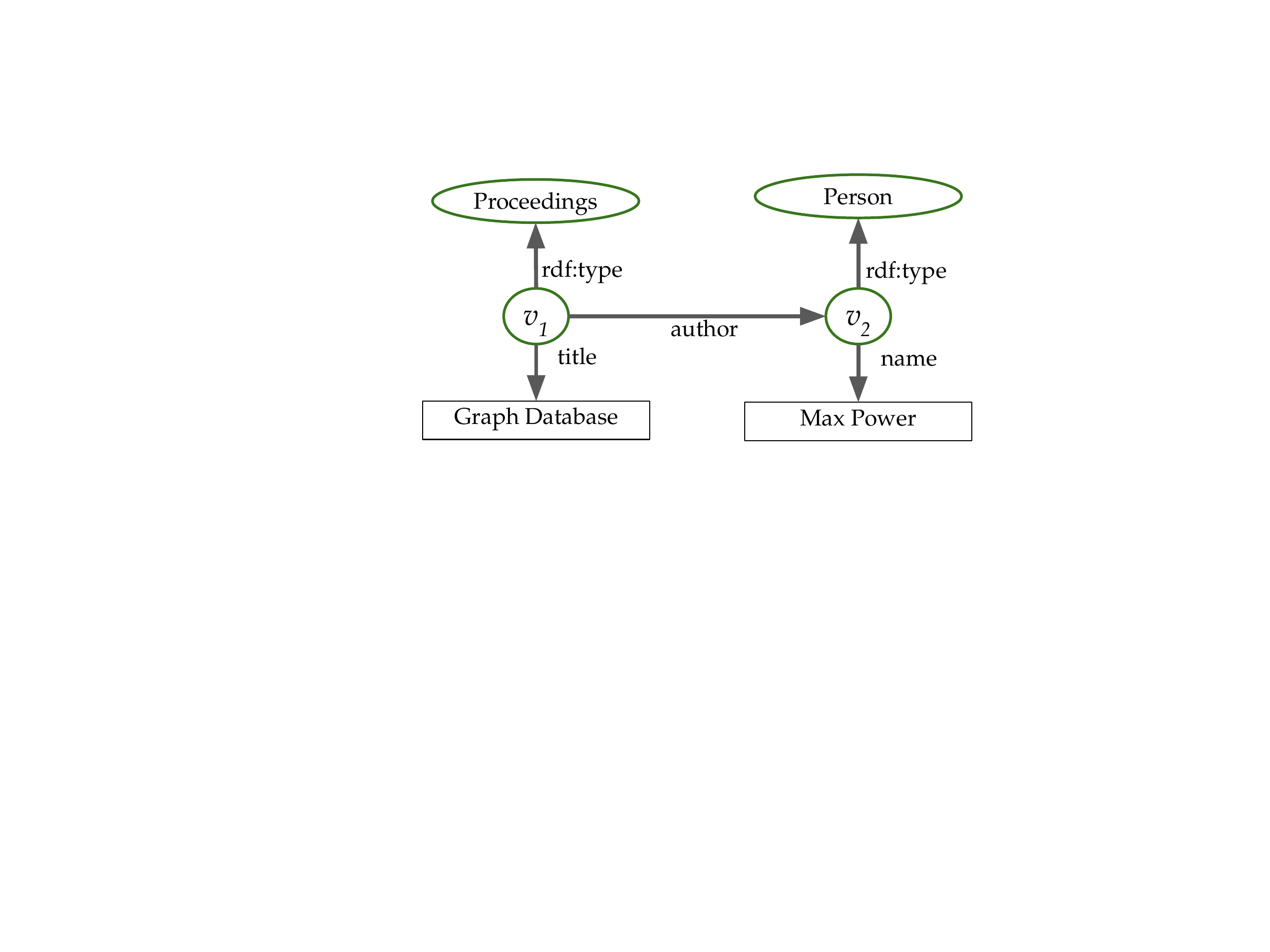}
\end{minipage}%
\caption{\label{fig:rdf-example-introduction} Simple RDF graph to demonstrate the relation between a set of triples (left) and its visualization as a graph (right).}
\end{figure}

%
An example RDF graph is shown in \cref{fig:rdf-example-introduction}.
On the left-hand side, the RDF graph is shown as set of triples.
On the right-hand side, it is depicted as a graph.
The vertex $v_1$ has the type set $\ts{v_1} = \lbrace \text{\texttt{Proceedings}}\rbrace$ and the property set $\ps{v_1} = \lbrace \text{\texttt{author, title}}\rbrace$.
The vertex $v_2$ has the type set $\ts{v_2} = \lbrace \text{\texttt{Person}}\rbrace$ and the property set $\ps{v_2} = \lbrace \text{\texttt{name}}\rbrace$.
But, $v_1$ has predicates $\lbrace \text{\texttt{author, title, rdf:type}}\rbrace$ and outgoing neighbors $\nsout{v_1} = \lbrace \texttt{Proceedings}, v_2, \text{\enquote{Graph Database}}\rbrace$.

A special characteristic of RDF graphs is their support for semantic labels, which allows the inference of implicit information. 
Such semantic labels are from ontologies, where semantic relationships between types and properties are denoted, \eg with predicates from the RDF Schema vocabulary.
RDF Schema (RDFS) and its entailment rules are standardized by the W3C~\cite{w3c-rdf-schema}.
A comprehensive overview of these rules is presented in~\cite{DBLP:books/daglib/0028543}.
For example, the semantics of a triple $(p, \texttt{rdfs:subPropertyOf}, p')\in G$ is that for any subject vertex~$s$ with $(s,p,o)\in G$, we can infer existence of the additional triple $(s,p',o)$.
This means, when using RDF Schema inference, each vertex may have more types and properties in its type set and property set, respectively. 

Finally, RDF supports the definition of named graphs~\cite{w3c-rdf}.
Following the W3C standard~\cite{w3c-rdf}, each named graph is a pair consisting of the graph name, which is an IRI, and an RDF graph as defined above.
Following Harth \etal~\cite{DBLP:conf/semweb/HarthUHD07}, we formalize named graphs by extending each triple $(s,p,o)$ to a tuple $((s, p, o), d)$.
The context of a tuple denotes the name of the data source from which the triple originated on the web~\cite{w3c-rdf-dataset-semantics}.
In this work, the context $d$ of a tuple is only used as payload, and not to determine the equivalence of vertices. 

\subsection{Semantic Structural Graph Summarization}
\label{sec:graph-summarization}
To compute a (semantic) structural \textbf{graph summary} $SG$ for a given data graph $G$, we partition the data graph into disjoint sets of vertices. 
In the case of semantic structural graph summaries, we partition the vertices based on equivalent subgraphs. 
In contrast to classic subgraph matching, we match only selected parts of the subgraphs, \eg match the type set and the property set of each vertex. 
We call the respective subgraphs containing the information necessary to determine the equivalence of two vertices the \textbf{schema structure} of the vertices. 
Which \textbf{features} of the input graph are considered to determine equivalent schema structures is defined by the \textbf{graph summary model}. 
%
For different tasks, different features of the summarized vertices are of interest, \eg the number of summarized vertices for cardinality computation or the data source (context $d$) for data search.
This information about the summarized vertices is called the \textbf{payload}.
Equivalence relations describe any graph partitioning in a formal way.
Thus, for \model{}, we define graph summarization via equivalence relations over vertices.

\begin{definition}
\label{def:eqrel}
An \textbf{equivalence relation} on a set $X$ (\eg a set of graph vertices) is a subset $\EQR \subseteq X \times X$ that is reflexive, symmetric, and transitive.
When $(x,y) \in \EQR$, we say that $x$ is equivalent to~$y$ and write $x \sim y$.
For any $y \in X$, the set $\{x\in X\mid x\sim y\}$ is called the \textbf{equivalence class of $y$}, typically denoted by $[y]_{\EQR}$. 
We denote by \enquote{$\taut$} the \textbf{tautology equivalence relation} $X\times X$, in which all elements of~$X$ are equivalent.
We denote by \enquote{$\id$} the \textbf{identity equivalence relation} $\{(x,x)\mid x\in X\}$, in which no two distinct elements of~$X$ are equivalent.
\end{definition}

\begin{remark}
For a given equivalence relation, any two equivalence classes either are disjoint or equal, so equivalence relations on a set~$X$ correspond precisely to partitions (decompositions) of~$X$~\cite{EncyMath}.
Furthermore, the intersection of two equivalence relations over~$X$ is also an equivalence relation.
\end{remark}


While the intersection of equivalence relations results in an equivalence relation, the same does not hold true for the union.
For example, consider the equivalence relations ${\sim_A} = \{(1,1), (2,2), (1,2), (2,1), (3,3)\}$ and ${\sim_B} = \{(1,1), (2,2), (3,3), (2,3), (3,2)\}$. 
The union of $\sim_A$ and~$\sim_B$ is not transitive, as it contains $(1,2)$ and $(2,3)$ but not $(1,3)$.
%
Thus, to combine equivalence relations in an \enquote{or-like} fashion to give a new equivalence relation, we define an extended union operator $\OR$.


\begin{definition}
\label{def:extended-union}
Let $\sim_{A}$ and~$ \sim_{B}$ be equivalence relations over the same set. 
We define the \textbf{extended union} ${\sim}_{A} \OR  {\sim}_{B}$ to be the transitive closure of $\sim_{A} \cup  \sim_{B}$.
That is, $\sim_{A} \OR \sim_{B}$ is the unique equivalence relation $\sim_\mathrm{ex}$ such that ${\sim}_{A} \cup {\sim}_{B} \subseteq {\sim}_\mathrm{ex}$ and for every equivalence relation~$\sim'$ such that $\sim_{A}\cup \sim_{B}\subseteq {\sim'}$, we have ${\sim}_{ex} \subseteq {\sim}'$.
\end{definition}

For \model{}, we define equivalence relations over the set of vertices appearing in the graph.
RDF graphs are defined as sets of triples $(s,p,o)$. 
However, since each triple has exactly one subject vertex $s$, an equivalence relation $\EQR_v$ over vertices induces an equivalence relation $\EQR_t$ over the triples, given by $((s,p,o),(s',p',o'))\in \EQR_t$ iff $(s,s')\in \EQR_v$.
Any partitioning of $G$'s vertices into disjoint subsets can be expressed as an equivalence relation over vertices in the data graph $G$.
Thus, we can describe graph summarization using an equivalence relation $\EQR$.
When two vertices are placed into the same partition, we say they are \textbf{summarized}.
The \textbf{graph summary} for $G$ with respect to~$\EQR$ is a labeled graph $SG$.
%
Each equivalence class $[y]_{\EQR}$ is represented in $SG$ in a so-called \textbf{vertex summary}.
The vertex summary $vs_y$ is a subgraph in $SG$ that is equivalent to the subgraphs of all summarized vertices $v \in [y]_{\EQR}$ in $G$ under $\EQR$.
For vertex summaries, we distinguish \textbf{primary vertices}, which are equivalence classes of~$\EQR$, and \textbf{secondary vertices}, which are equivalence classes of the relations from which $\EQR$~is defined~\cite{CIKM-incremental-summary}.
Furthermore, we can attach \textbf{payload}~\cite{DBLP:conf/kcap/GottronSKP13} to each primary vertex in a vertex summary $vs \subseteq SG$.
The payload can be tailored for different purposes, \eg to contain the number of summarized vertices.
For \model{}, we define a set of \textbf{payload elements} $\PAY$ to implement a specific task.
Payload elements map vertex summaries to payload.
%
%
Formally, we can now denote a (semantic) structural graph summary model as a $3$-tuple of a data graph $G$, an equivalence relation $\EQR$, and a set of payload elements $\PAY$.
\begin{definition}
\label{def:index}
A \textbf{structural graph summary model} is a tuple $(G, \EQR, \PAY)$, where $G$~is the data graph, $\EQR$ is an equivalence relation over vertices in~$G$, and $\PAY$ is a set of payload elements.
\end{definition}

\section{Analysis of Existing Semantic Structural Graph Summary Models}
\label{sec:related-work}
Throughout the literature, different attempts have been made to classify and group the different graph summarization approaches~\cite{DBLP:journals/csur/LiuSDK18,DBLP:journals/corr/abs-2004-14794,DBLP:journals/pvldb/KhanBB17,DBLP:journals/vldb/CebiricGKKMTZ19}.
%
In this work, we focus on structural graph summaries, \ie graph summaries that precisely capture the structure of the input graph~\cite{DBLP:journals/vldb/CebiricGKKMTZ19}.
This excludes in particular statistical approaches that generate approximate data descriptions~\cite{DBLP:journals/vldb/CebiricGKKMTZ19}.
As pointed out by Fan \etal~\cite{DBLP:conf/sigmod/FanLLTWW11}, many real-life applications require exact matches.
Still, there exists a large variety of exact (semantic) structural graph summaries~\cite{DBLP:journals/vldb/CebiricGKKMTZ19}.


In the following, we analyze in detail existing (semantic) structural graph summaries with respect to the captured schema structure, \ie what features of the input graph are used to summarize vertices.
Existing surveys about graph summaries cover a wider range of approaches and, thus, lack this level of granularity~\cite{DBLP:journals/csur/LiuSDK18,DBLP:journals/corr/abs-2004-14794,DBLP:journals/pvldb/KhanBB17,DBLP:journals/vldb/CebiricGKKMTZ19}. 
\cref{tab:index-model-features} shows a cross-table of each analyzed structural graph summary model and its features.
In total, we analyzed $19$ graph summary models and identified $12$ different features.
We organize the structural graph summary models and the features into different groups. 
We distinguish features that only use triple information (triple features), features that define how features of multiple vertices are combined (subgraph features), and features that define explicit semantic rules such joining and inference (semantic rule features).
%
%
Each group of features adds another level of complexity, \ie intuitively, the computational complexity grows when features of different groups are used. 
%
%
As one can see, there is no single graph summary model that supports all features.
However, we see common combinations of features.
In the following, we discuss the graph summary models along the identified features shown in \cref{tab:index-model-features} from left to right.

\begin{table*}[!t]
\caption{\label{tab:index-model-features}Structural graph summary models found in the literature and what features they use (X) and do not use (-) to capture the schema structure of vertices. The features are grouped by features that use triple information only (triple features), features that define how features of multiple vertices are combined ({subgraph features}), and features that define explicit semantic rules ({semantic rule features}). 
}
\setlength{\tabcolsep}{5pt}
\small
\begin{tabularx}{\linewidth}{p{0.3cm} p{5.1cm} *{4}{p{0.5cm}}| *{4}{p{0.5cm}}| *{4}{p{0.5cm}}}

\toprule
& & \multicolumn{4}{c|}{\textit{Triple features}} & \multicolumn{4}{c|}{\textit{Subgraph features}} & \multicolumn{4}{c}{\textit{Semantic rule features}}\\
& \backslashbox{\textbf{Graph summary}}{\textbf{Feature}} & \rotateHeading{Property sets} & \rotateHeading{Type sets} &  \rotateHeading{Label sets} & \rotateHeading{Neighbor vertex ID} & \rotateHeading{Neighbor triple} & \rotateHeading{Predicate path} & \rotateHeading{($\boldsymbol{k}$-)bisimulation}& \rotateHeading{Incoming property sets} & \rotateHeading{OR combination} & \rotateHeading{Related properties}  & \rotateHeading{RDF Schema} & \rotateHeading{OWL SameAs} \\

\arrayrulecolor{black!100}\midrule

\parbox[t]{4mm}{\multirow{4}{*}{\rotatebox[origin=c]{90}{\textit{Simple}}}}
& Attribute-based Collection~\cite{DBLP:conf/dexaw/CampinasPCDT12} & X & - & - & - & - & - & - & - & - & - & - & -\\
& Class-based Collection~\cite{DBLP:conf/dexaw/CampinasPCDT12}& - & X & - & - & - & - & - & - & - & - & - & -\\
& Characteristic Sets~\cite{DBLP:conf/icde/NeumannM11} & X & - & - & - & - & - & -& X & - & - & - & - \\
& SemSets~\cite{DBLP:conf/www/CiglanNH12} & - & - & - & X & - & X & - & - & - & - & - & -\\

\arrayrulecolor{black!50}\midrule
\arrayrulecolor{black!100}

\parbox[t]{4mm}{\multirow{14}{*}{\rotatebox[origin=c]{90}{\textit{Complex}}}}
& SchemEX~\cite{DBLP:journals/ws/KonrathGSS12} & X & X & - & - & X & X & - & - & - & - & - & -\\

& SchemEX+U+I~\cite{DBLP:conf/dexa/BlumeS20} & X & X & - & - & X & X & - & - & - & - & X & X\\

& ABSTAT~\cite{DBLP:conf/esws/SpahiuPPRM16a} & X & X & - & - & X & X & - & - & - & - & (X) & -\\

& LODex~\cite{lodex2015} & X & X & - & - & X & X & - & - & - & - & - & -\\

& Loupe~\cite{loupe2015} & X & X & - & - & X & X & - & - & - & - & - & -\\

& TermPicker~\cite{DBLP:conf/esws/SchaibleGS16} & X & X & - & - & X & - & - & - & - & - & - & -\\

\arrayrulecolor{black!20}\cline{2-14}
\arrayrulecolor{black!100}
%
& Weak Summary~\cite{DBLP:journals/vldb/GoasdoueGM20} &X &- &- &- &X &X &X &X &X &X &X & -\\

& Strong Summary~\cite{DBLP:journals/vldb/GoasdoueGM20} &X &- &- &- &X &X &X &X &- &X &X & -\\

& Typed Weak Summary~\cite{DBLP:journals/vldb/GoasdoueGM20} &X &X &- & - &X &X &X &X &X &X &X & -\\

& Typed Strong Summary~\cite{DBLP:journals/vldb/GoasdoueGM20} &X&X &- & - &X &X &X &X &- &X &X & -\\

\arrayrulecolor{black!20}\cline{2-14}
\arrayrulecolor{black!100}

& Tran \etal~\cite{DBLP:journals/tkde/TranLR13} &X &- &X &- &X &X &X &- &- &- & - & -\\

& A($k$)-index~\cite{DBLP:conf/icde/KaushikSBG02} &X &- &- &- &X &X &X &- &- &- & - & - \\

& T-index~\cite{DBLP:conf/icdt/MiloS99} &- &- &- &- &X &X &X &X &- &- &- & -\\

& Consens \etal~\cite{DBLP:journals/pvldb/ConsensFKP15}  &X &X &- &- &X &X &X &- &- &- &- & - \\

& Schätzle \etal~\cite{DBLP:conf/sigmod/SchatzleNLP13} &X &- &- &X &- &X &X &- &- &- &- & -\\

\arrayrulecolor{black!100}\bottomrule

\end{tabularx}
\end{table*}

\subsection{Triple Features}
\label{sec:triple-features}
Triple features are solely based on outgoing triples of vertices.
To compute the equivalence of two vertices $s$ and~$s'$, we only compare triples where the subject is $s$ or~$s'$.

\paragraph{Property sets}
The most commonly used feature in structural graph summaries is using properties to compute the schema of vertices.
More specifically, for each vertex $s$ in the data graph the property set $\ps{s}$ is compared.
Campinas \etal~\cite{DBLP:conf/dexaw/CampinasPCDT12} proposed so-called \enquote{Attribute-based Collections}, a graph summary that relies solely on property sets to compute the schema structure of vertices.
If two vertices $s,s'$ share the same property set, \ie $\ps{s} = \ps{s'}$, they are considered equivalent, thus, are summarized together.

\paragraph{Type sets}
Another commonly used feature is using the vertices' types to compute the schema.
Here, for each vertex $s$ in the data graph, the type set $\ts{s}$ is compared.
If two vertices $s,s'$ share the same type set, \ie $\ts{s} = \ts{s'}$, they are considered equivalent.
Campinas \etal~\cite{DBLP:conf/dexaw/CampinasPCDT12} proposed a second graph summary, called \enquote{Class-based Collections}, which uses only the vertices' type sets to compute the schema.
Both graph summaries, Attribute-based Collections and Class-based Collections, were developed to enhance SPARQL query formulations by providing recommendations~\cite{DBLP:conf/dexaw/CampinasPCDT12}.

\paragraph{Label sets}
Tran \etal~\cite{DBLP:journals/tkde/TranLR13} proposed the feature of label parameterization for graph summaries.
With the label parameterization, only a subset of all edge labels are used to compute the schema.
More precisely, one defines a set of predicates $P_l$, the so-called label set, which are ignored when determining the equivalence of vertices.
Tran \etal's graph summary combines property sets $\ps{s}$ with label sets.
Furthermore, they combine this with k-bisimulation (see below). 

\paragraph{Neighbor vertex ID}
The final triple feature is using the identity of outgoing neighbors $\nsout{s}$ to determine the equivalence of vertices.
It appears that no existing graph summary summarizes vertices solely by comparing the neighbor identities. 
However, SemSets~\cite{DBLP:conf/www/CiglanNH12} summarize vertices that share the same outgoing predicates, which are linked to the same vertices.
To check if two vertices $s,s'$ are equivalent under SemSets, all triples where $s$ or~$s'$ are the subject vertices are compared.
For each triple $(s,p,o)  \in G$ there has to be a triple $(s',p,o) \in G$, and vice versa.
Thus, they combine neighbor vertex identifiers $\nsout{s}$ with predicate paths (see below).  
SemSets were developed to discover semantically similar sets of vertices in graphs and to use these vertex sets to improve keyword-based ad-hoc retrieval.

\subsection{Subgraph Features}
\label{sec:sub-graph-features}
The neighbor vertex identifier is the most direct approach to incorporate neighbor information that leads to a wider range of summary models that consider neighbor information, \eg vertices in $\nsout{s}$.
We classify features as subgraph features when they combine triple features of multiple vertices. 

\paragraph{Neighbor triple}
SchemEX~\cite{DBLP:journals/ws/KonrathGSS12}, SchemEX+U+I~\cite{DBLP:conf/dexa/BlumeS20}, ABSTAT~\cite{DBLP:conf/esws/SpahiuPPRM16a}, LODeX~\cite{lodex2015}, and Loupe~\cite{loupe2015} summarize vertices $s$ and~$s'$ based on a common type set and common properties linking to vertices with the same type sets.
This means, in order to compute the schema of one vertex~$s$, also the type sets of outgoing neighbors $\nsout{s}$ are required to be equivalent, \ie we compare neighbor triples.
In contrast to SemSets~\cite{DBLP:conf/www/CiglanNH12}, these approaches do not use the neighbor vertex identifiers $\nsout{s}$ but, instead, use the type set $\ts{o}$ for each $o \in \nsout{s}$.
SchemEX~\cite{DBLP:journals/ws/KonrathGSS12}, SchemEX+U+I~\cite{DBLP:conf/dexa/BlumeS20}, ABSTAT~\cite{DBLP:conf/esws/SpahiuPPRM16a}, LODeX~\cite{lodex2015}, and Loupe~\cite{loupe2015} combine type sets $\ts{s}$, property sets $\ps{s}$, and neighbor type sets $\ts{o}$ using predicate paths (see below).
The summary models were developed for data source search and exploration. 

\paragraph{Predicate path}
As indicated in the previous two features, almost all analyzed graph summaries that use neighbor information combine the schema structures using predicate paths, \ie they compare which predicates link to which neighbors.
A predicate path compares via which path of predicates a vertex is connected to neighboring vertices' schema structures.  
For example, SchemEX~\cite{DBLP:journals/ws/KonrathGSS12}, SchemEX+U+I~\cite{DBLP:conf/dexa/BlumeS20}, ABSTAT~\cite{DBLP:conf/esws/SpahiuPPRM16a}, LODeX~\cite{lodex2015}, and Loupe~\cite{loupe2015} consider which property links to which type set.
TermPicker~\cite{DBLP:conf/esws/SchaibleGS16} follows a different strategy to integrate the schema of neighboring vertices. 
TermPicker summarizes vertices~$s$ based on having the same type set $\ts{s}$, the same property set $\ps{s}$, and the same type set $\ts{o}$ of all $o \in \nsout{s}$.
Consequently, TermPicker's graph summaries compress all type sets of all neighbors into a single type set.
Thus, TermPicker's graph summaries do not contain information about which \emph{specific} property links to which neighbor. 

\paragraph{($k$-)bisimulation}
Many graph summaries compute the schema of vertices by taking into account the schema of neighbors over multiple hops~\cite{DBLP:journals/ws/KonrathGSS12,DBLP:conf/sigmod/QunLO03,DBLP:journals/tkde/TranLR13,DBLP:conf/icde/KaushikSBG02}.
This is commonly defined as a bisimulation.
Bisimulation operates on state transition systems and defines an equivalence relation over states~\cite{Bisimulation:2009}.
Two states are equivalent (or bisimilar) if they change into equivalent states with the same type of transition.
Interpreting a labeled graph as a representation of a state transition system allows us to apply bisimulation on graph data to discover structurally equivalent parts.

In practice, graph summaries usually define a stratified $k$-bisimulation~\cite{DBLP:journals/ws/KonrathGSS12,DBLP:conf/sigmod/QunLO03,DBLP:journals/tkde/TranLR13,DBLP:conf/icde/KaushikSBG02}.
A stratified bisimulation restricts the maximum path length to $k$~edges in the connected subgraph.
This increases the chance that two vertices are considered equivalent.
An efficient $k$-bisimulation algorithm was proposed by Kaushik \etal~\cite{DBLP:conf/icde/KaushikSBG02} but there are also efficient implementations specialized to their corresponding graph summary, \eg by Konrath \etal~\cite{DBLP:journals/ws/KonrathGSS12} and Goasdou\'e \etal~\cite{DBLP:journals/vldb/GoasdoueGM20}.
Tran \etal~\cite{DBLP:journals/tkde/TranLR13} propose, in addition to the label parameterization feature described above, also the height parameterization feature, formulating $k$-bisimulation as feature for graph summaries. 

Some graph summaries combine the feature of using only incoming or only outgoing properties with the $k$-bisimulation feature~\cite{DBLP:conf/icdt/MiloS99,DBLP:journals/pvldb/ConsensFKP15,DBLP:conf/sigmod/SchatzleNLP13}.
This is referred to as backward $k$-bisimulation and forward $k$-bisimulation, respectively~\cite{DBLP:journals/vldb/GoasdoueGM20}. 
Milo and Suciu~\cite{DBLP:conf/icdt/MiloS99} developed the so-called T-index to support path queries in semi-structured databases.
This summarizes vertices~$s$ based on the common set of incoming property-paths, \ie they use $k$-bisimulation only on incoming property sets $\psin{s}$.
Consens \etal~\cite{DBLP:journals/pvldb/ConsensFKP15} propose a structural graph summary model to support navigational SPARQL queries, so-called Extended Property Paths (EPPs).
They summarize vertices~$s$ based on the common set of outgoing property-paths, \ie they use $k$-bisimulation only on outgoing property sets $\ps{s}$.
In addition, for each hop, the type sets $\ts{s}$ have to be equivalent.
Schätzle \etal~\cite{DBLP:conf/sigmod/SchatzleNLP13} developed a similar approach like Consens \etal~\cite{DBLP:journals/pvldb/ConsensFKP15}.
The difference is that Schätzle \etal{} do not consider type sets, but object equivalences.
Furthermore, they do not distinguish between types and properties. 
Thus, they use all predicates. 
This means, for each hop the same vertex is connected over the same predicate (neighbor vertex ID feature).
For $k=1$, Schätzle \etal's summary model is equivalent to the SemSets~\cite{DBLP:conf/www/CiglanNH12} model.

\paragraph{Incoming property sets}
Incoming property sets are frequently used in combination with $k$-bisimulation.
To the best of our knowledge, there is no graph summary model that summarizes solely based on incoming property sets $\psin{s}$.
But, Characteristic Sets~\cite{DBLP:conf/icde/NeumannM11} summarize two vertices $s, s'$ that have the same outgoing property sets $\ps{s} = \ps{s'}$ and the same incoming property sets $\psin{s} = \psin{s'}$.
Characteristic Sets were designed for cardinality estimations of queries in RDF databases.
Analogously, Goasdou\'e \etal~\cite{DBLP:journals/vldb/GoasdoueGM20} define the Strong Summary.
The Strong Summary summarizes vertices $s$ and~$s'$ if they have the same property sets $\ps{s} = \ps{s'}$ and the same incoming property set $\psin{s} = \psin{s'}$.
Furthermore, they propose the Typed Strong Summary, which summarizes vertices $s$ and~$s'$ based on two conditions: (1) if they have empty type sets $\ts{s} = \ts{s'} = \emptyset$ and they have the same property sets $\ps{s} = \ps{s'}$ and the same incoming property set $\psin{s} = \psin{s'}$ or (2) if they have the same non-empty type sets $\ts{s} = \ts{s'}\neq\emptyset$~\cite{DBLP:journals/vldb/GoasdoueGM20}.
Both the Strong Summary and the Typed Strong Summary also allow $k$-bisimulation and semantic rule features such as taking so-called \enquote{related properties} into account and exploit RDF Schema inferencing (see below).

\subsection{Semantic Rule Features in Graph Summaries}
\label{sec:rule-features}
The last group includes features that define explicit (semantic) rules. 

\paragraph{OR combination}
Goasdou\'e \etal~\cite{DBLP:journals/vldb/GoasdoueGM20} define the Weak Summary using OR combination (see extended union $\OR$, \cref{def:extended-union}).
In the Weak Summary, two vertices $s$ and~$s'$ are equivalent if they have the same property set $\ps{s} = \ps{s'}$ or the same incoming property set $\psin{s} = \psin{s'}$ (or both).
Analogously to the Typed Strong Summary, they define the Typed Weak Summary, which relaxes the first condition of the equivalence of the property set and the incoming property set using the OR combination of the Weak Summary~\cite{DBLP:journals/vldb/GoasdoueGM20}.

\paragraph{Related properties}
Goasdou\'e \etal~\cite{DBLP:journals/vldb/GoasdoueGM20} also propose to include property relations.
Two properties $p$ and~$p'$ are source-related if they co-occur in any property set $\ps{s}$ of any vertex~$s$ and they are target-related if they co-occur in any incoming property set $\psin{s}$ of any vertex~$s$.
They developed this feature to generate comprehensible graph visualizations. 

\paragraph{RDF Schema}
Several semantic structural graph summaries use RDF Schema inferencing to enhance their summaries.
ABSTAT~\cite{DBLP:conf/esws/SpahiuPPRM16a} exploits RDF Schema type hierarchies to compute so-called minimal patterns.
They select the minimal number of types, \ie they only keep the most specific types from the RDF Schema type hierarchy.
Goasdou\'e \etal~\cite{DBLP:journals/vldb/GoasdoueGM20} exploit RDF Schema type hierarchies, property hierarchies, and RDF Schema domain and RDF Schema range.
With domain and range, types for the subject vertex and the object vertex can be inferred.
They also propose a so-called shortcut for inferencing, which improved the time needed to perform the inferencing in graph summaries by up to $94\%$~\cite{DBLP:journals/vldb/GoasdoueGM20}.

\paragraph{OWL SameAs}
SchemEX+U+I~\cite{DBLP:conf/dexa/BlumeS20} also uses the full RDF Schema inferencing but also exploits the semantics of the \texttt{owl:sameAs} property. 
This property is part of W3C's Web Ontology Language (OWL)~\cite{w3c-owl}, which is heavily used in the context of RDF graphs.
The \texttt{owl:sameAs} property defines an equivalence relation~\cite[Section 4.2]{w3c-owl}, intended to identify vertices that represent the same real-world entity.
To compute the schema structure of one vertex $v$, the schema structures of all vertices $v'$ in the weakly connected components in an \texttt{owl:sameAs}-labeled subgraph of $G$ are merged (see Ding \etal~\cite{DBLP:conf/semweb/DingSSM10} for details on \texttt{owl:sameAs} networks).
SchemEX+U+I was developed for a data search task. 

\subsection{Summary}
There exists a large variety of structural graph summary models that use different combinations of features. 
However, none of the existing approaches covers all features. 
In the past, the idea of a single, generalizable framework to compute structural graph summaries has been discussed a few times. 
To the best of our knowledge, the idea of using equivalence relations to define structural graph summaries, \ie to partition vertices of a graph, has been proposed before~\cite{DBLP:conf/sigmod/QunLO03,DBLP:journals/tkde/TranLR13,DBLP:conf/semweb/CebiricGM17}. 
In particular, the concept of bisimulation has been used to define a single, adaptive framework for graph summaries based on equivalence relations~\cite{DBLP:conf/sigmod/QunLO03, DBLP:journals/tkde/TranLR13}.
However, none of the existing approaches define a language to define these equivalence relations, which also allows a single, parameterized algorithm to compute them.
The proposed features and the corresponding algorithms only implement a subset of features, as shown in \cref{tab:index-model-features}.
Thus, a language for semantic structured graph summarizes is needed that incorporates all of the different features and that allows us to flexibly define (semantic) structural graph summaries.
In the next section, we define our flexible language called \model{} and demonstrate how it covers all features found in the literature.


\section{Definition of \model{}}
\label{sec:formal-defintions}
We formally define the elements and parameterizations of \model{}.
An overview is given in \cref{tab:building-blocks}.
First, we define three different simple schema elements in \cref{sec:schema-elements}.
Based on simple schema elements, we define complex schema elements.
Second, we define in \cref{sec:parameters} six parameterizations to further specialize the schema elements.
Third, we define three payload elements in \cref{sec:payload-formal}.
Finally, in \cref{sec:modeling-with-fluid}, we show how all graph summary models analyzed in \cref{sec:related-work} are defined with \model{}.

\begin{table*}[!t]
\caption{Overview of \model 's schema elements and parameterizations.}
{\renewcommand\arraystretch{1.3}
\small
\setlength{\tabcolsep}{8pt}
\begin{tabularx}{\textwidth}{>{\raggedright\arraybackslash}p{5.5cm} >{\raggedright\arraybackslash}p{6.75cm} >{\raggedright\arraybackslash}p{2.6cm}}
\toprule

\textbf{Schema Elements (SE)} & \textbf{Description} & \textbf{Details} \\

\midrule

Simple Schema Element (SSE) & Triple-based summarization of vertices &  \cref{def:poc,def:oc,def:pc}, \cref{fig:example-predicate-object-cluster,fig:example-predicate-cluster,fig:example-object-cluster}\\

Complex Schema Element (CSE) & Summarization of vertices using combinations of Schema Elements (SEs) & \cref{def:complex-schema-element}, \cref{fig:example-cse-2} \\

\toprule
\textbf{Parameterizations} & \textbf{Description} & \textbf{Details}  \\
\midrule

Label parameterization $lp(SSE,P_l)$ (short: $SSE_{P_l}$) & SSE ignores existence of predicates not in $P_l$& \cref{def:labelParameterization} \\

Set parameterization $sp(SSE,S)$ (short: $SSE|_{S}$) & Application of SSE is restricted to vertices containing only specified predicates and/or objects & \cref{def:set-parameterization} \\
Chaining parameterization $cp(CSE,k)$ (short: $CSE^k$) & Recursively apply CSE for each connected vertex up to $k$ hops& \cref{def:chaining-parameter}\\

Direction parameter $dp(SE,\delta)$\hspace{5mm} (short: $\delta\mhyphen SE$)& Summarize vertices following the direction parameter $\delta = \{\outdir,\indir, \bidir \}$ based on outgoing predicates ($\outdir$), incoming predicates ($\indir$), or outgoing and incoming predicates ($\bidir$)&\cref{def:direction-param}  \\

Inference parameterization $op(G, VG)$ (short: $G_{VG}$)& Include ontology reasoning by inferring additional triples from a vocabulary graph $VG$, \eg using $\VGRDFS$ & \cref{def:inference-param} \\

Instance parameter $ip(SE, \Delta)$ \hspace{10mm} (short: $SE[\Delta]$)& Unions of explicitly equivalent vertices & \cref{def:instance-param} \\

\bottomrule
\end{tabularx}
\label{tab:building-blocks}
}
\end{table*}

\subsection{Schema Elements}
\label{sec:schema-elements}
We distinguish simple schema elements and complex schema elements.
Simple schema elements enable triple-based vertex summarization (compare triple features in \cref{tab:index-model-features}).
Intuitively, this means they summarize vertices without considering any kind of neighbor information, \ie they capture the \enquote{local} schema of vertices.
Recall from \cref{sec:graph-model} that we assume the graph to be defined using triples of subject vertex, edge label (predicate), and object vertex, \ie $(s, p, o)$.
Local in this sense means that equivalence of two vertices $s,s'$ depends only on the edges where $s$ or $s'$ is the source vertex, \ie the subject of the triples. 
In contrast, complex schema elements allow us to capture the schema beyond the scope of a single vertex (compare subgraph features in \cref{tab:index-model-features}).
In the following, we formally define three simple schema elements as well as the complex schema element.

\subsubsection{Simple Schema Elements}
\label{sec:simple-schema-elems}
\begin{figure}[t]
    \centering
    \begin{subfigure}[t]{0.48\textwidth}
        \centering
        \includegraphics[width=0.9\textwidth,trim=7cm 10cm 5cm 3cm, clip=true]{\images RDFGraph}
        \caption{\label{fig:example-data-graph} Example RDF data graph as introduced in \cref{fig:rdf-example-introduction}.}
    \end{subfigure}%
    \quad
    \begin{subfigure}[t]{0.48\textwidth}
        \centering
        \includegraphics[width=0.9\textwidth,trim=4cm 10cm 8cm 3cm, clip=true]{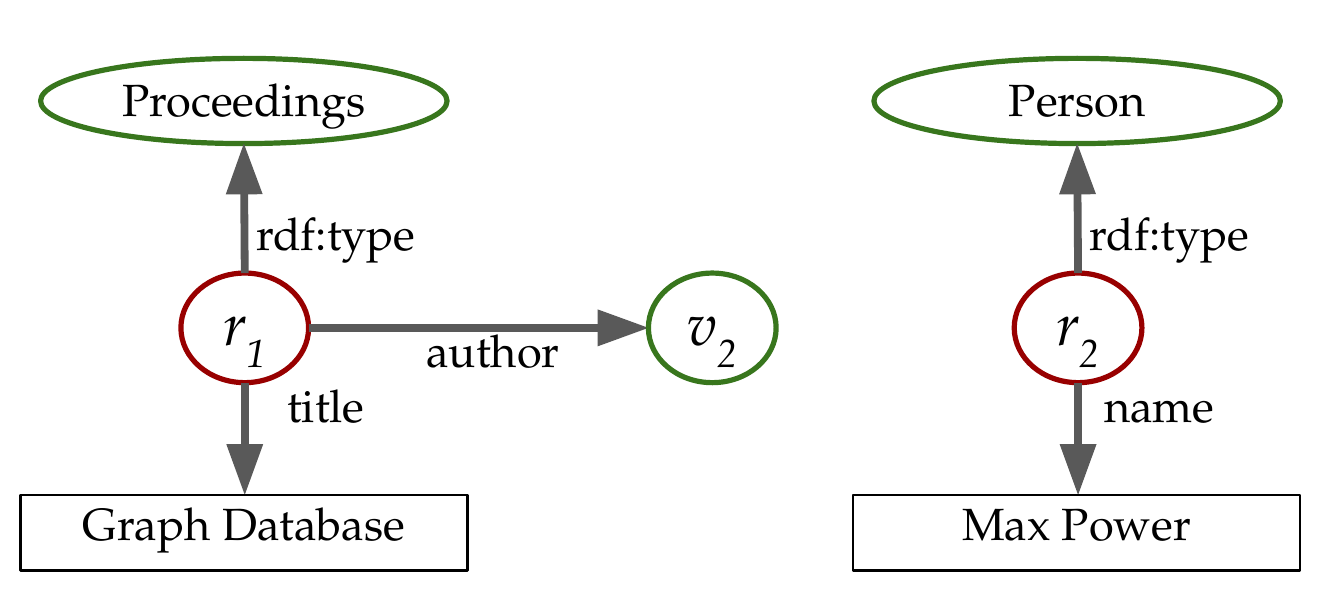}
        \caption{\label{fig:example-predicate-object-cluster} The two vertex summaries that summarize $v_1$ and $v_2$ according to the $\POC$ equivalence relation.}
    \end{subfigure}
    \quad
    \begin{subfigure}[t]{0.48\textwidth}
        \centering
        \includegraphics[width=0.9\textwidth,trim=4cm 10cm 8cm 3cm, clip=true]{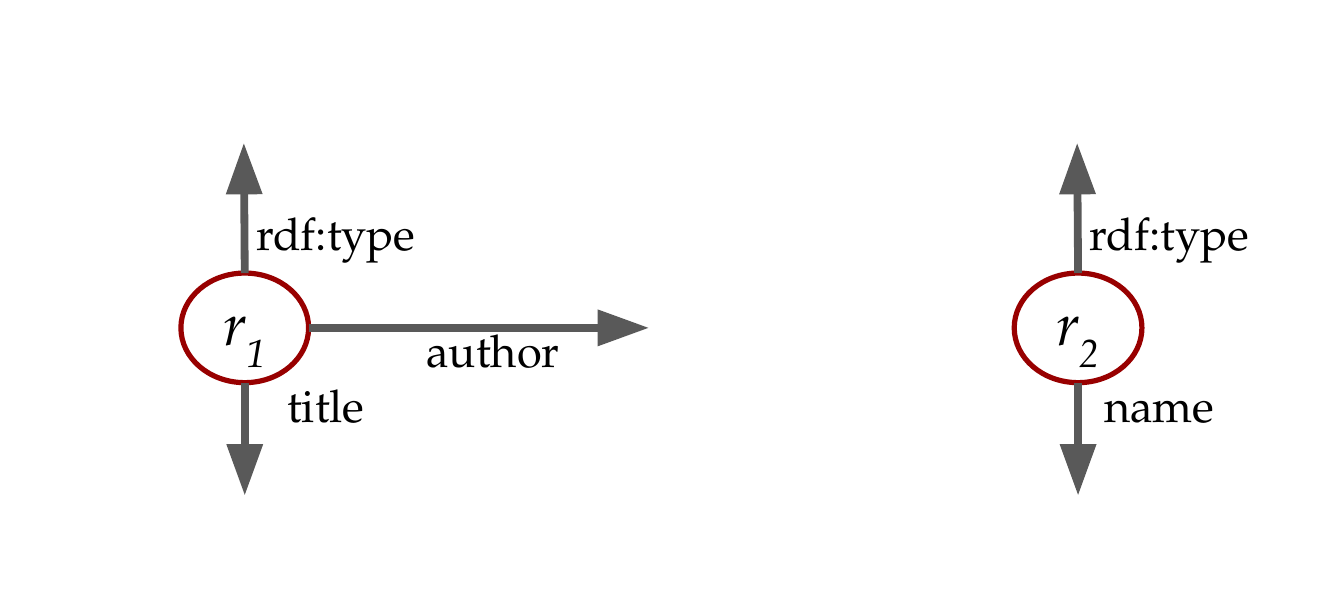}
        \caption{\label{fig:example-predicate-cluster} The two vertex summaries that summarize $v_1$ and $v_2$ according to the $\PC$ equivalence relation.}
    \end{subfigure}%
    \quad
    \begin{subfigure}[t]{0.48\textwidth}
        \centering
        \includegraphics[width=0.9\textwidth,trim=4cm 10cm 8cm 3cm, clip=true]{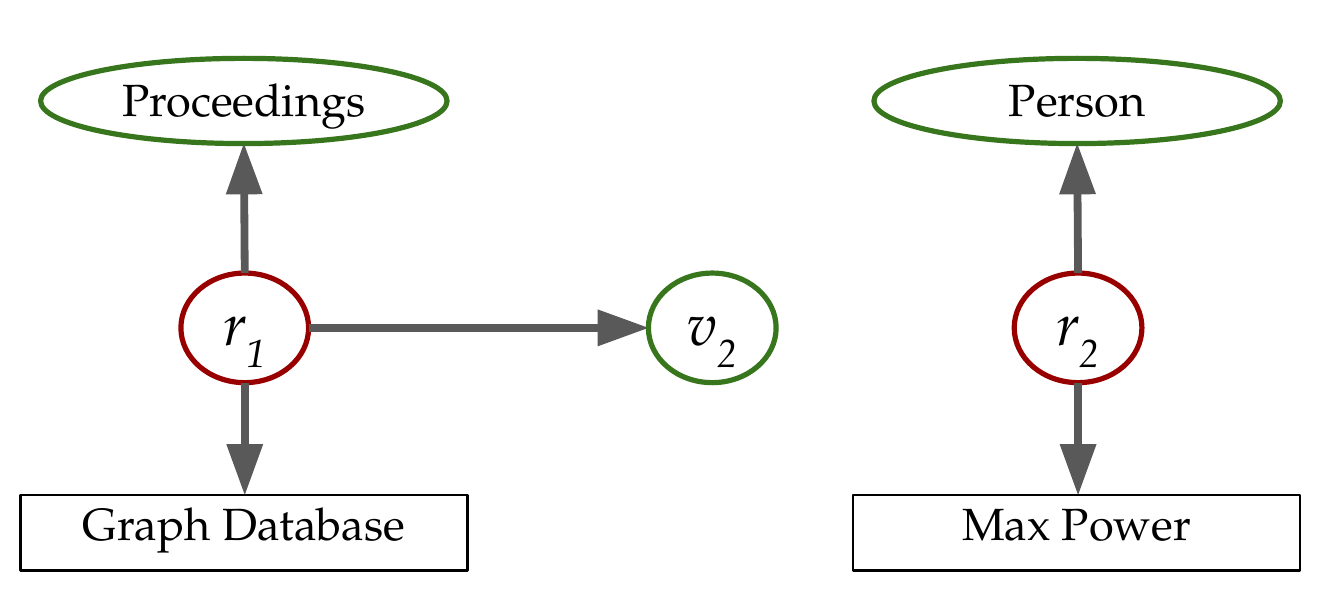}
        \caption{\label{fig:example-object-cluster} The two vertex summaries that summarize $v_1$ and $v_2$ according to the $\OC$ equivalence relation.}
    \end{subfigure}%
    \caption{\label{fig:fluid-example-figures} An example data graph (a) and the vertex summaries in the graph summary $SG$ that summarize vertices $v_1$ and $v_2$ using $\POC$ (b), $\PC$ (c), and $\OC$ (d). The vertex summary identified by the primary vertex $r_1$ summarizes $v_1$ and the vertex summary identified by the primary vertex $r_2$ summarizes $v_2$. Both vertex summaries are the subgraphs starting from their primary vertex (see \cref{sec:graph-summarization}).}
\end{figure}

Simple schema elements summarize vertices $s,s'$ by comparing all triples $(s,p,o) \in G$ and all triples $(s',p',o') \in G$.
%
We define three simple schema elements: predicate-object cluster ($\POC$), predicate cluster ($\PC$), and object cluster ($\OC$).
Each simple schema element compares vertices following a different strategy, \ie compare only the predicates, only the objects, or both (see analysis of triple features in \cref{sec:triple-features}).
\cref{fig:fluid-example-figures} gives an example RDF graph and the corresponding vertex summaries using simple schema elements.

\begin{definition}
\label{def:poc}
The \textbf{Predicate--Object Cluster} ($\POC$) partitions the data graph by comparing the triples based on common predicates linking to common objects.
For two vertices $s, s'$, the equivalence relation $\POC$ holds true, iff for each triple $(s,p,o) \in G$ there is a triple $(s',p,o)\in G$, and vice versa.
\end{definition}

\cref{fig:example-predicate-object-cluster} shows two vertex summaries that summarize $v_1$ and $v_2$, respectively, following the predicate-object cluster $\POC$ equivalence.

\begin{definition}
\label{def:pc}
    The \textbf{Predicate Cluster} ($\PC$) partitions the data graph by common predicates of vertices: $(s,s')\in \PC$ iff, for each triple $(s, p, o)\in G$, there is a triple $(s', p, o')\in G$ for some~$o'$, and vice versa.
\end{definition}

Note that the predicate cluster is not equivalent to saying $\ell_P(s)=\ell_P(s')$ due to the special treatment of the RDF predicate $\texttt{rdf:type}$ (see \cref{sec:graph-model}).
As shown in \cref{fig:example-predicate-cluster}, the predicate cluster includes the \texttt{rdf:type} predicate.
However, the \texttt{rdf:type} predicate is explicitly excluded from property sets. 
Rather, \texttt{rdf:type} defines type sets (see again \cref{sec:graph-model}).

\begin{definition}
\label{def:oc}
    The \textbf{Object Cluster} ($\OC$) partitions the data graph by common objects of vertices: $(s,s')\in \OC$ iff, for each triple $(s, p, o)\in G$, there is a triple $(s', p', o)\in G$ for some~$p'$, and vice versa.
\end{definition}

Note that $\POC\neq\PC\cap\OC$. 
Two instances are equivalent under $\PC\cap\OC$ if they contain the same predicates and the same objects, whereas $\POC$ requires the same predicate--object pairs. 
For example, consider the novel \emph{David Copperfield} by Charles Dickens and a hypothetical biography \emph{Charles Dickens} by the illusionist David Copperfield. 
These two books are equivalent under $\PC\cap\OC$ because each has properties ``author'' and ``title'' and each has objects ``David Copperfield'' and ``Charles Dickens''. However, they are not equivalent under $\POC$ because, \eg one has ``Charles Dickens'' as author and the other does not.
This acknowledges the predicate path feature in \cref{tab:index-model-features}.

%
%

\subsubsection{Complex Schema Elements}
\label{sec:complex-elements}
\model's three simple schema elements summarize vertices by comparing outgoing triples.
However, we also need to support schema structures that go beyond the scope of a single vertex.
Thus, we define the complex schema element as a generalization of a simple schema element.
The simple schema elements are combinations of equivalence relations by using the identity equivalence $\id$ on properties and/or objects.
Complex schema elements extend on this and allow arbitrary equivalence relations over subjects, properties, and objects.
Thus, they can be considered as templates to combine any number of (simple) schema elements.
This allows, \eg applying equivalence relations defined by simple schema elements on objects, \ie use the local schema of neighboring vertices.

\begin{definition}
\label{def:complex-schema-element}
A \textbf{complex schema element} is a $3$-tuple $CSE := (\sim^{s}$, $\sim^{p}$, $\sim^{o})$, where  $\sim^{s}$, $\sim^{p}$, and~$\sim^{o}$ are equivalence relations.
Two vertices $s, s'$ are equivalent under this CSE iff, for every triple $(s, p, o)\in G$, there is a triple $(s', p', o')\in G$ such that $s \sim^{s} s'$, $p \sim^{p} p'$, and $o \sim^{o} o'$, and vice versa.
\end{definition}

The following examples show how to combine equivalence relations using complex schema elements.
This allows us to incorporate the schema of neighboring vertices $\Gamma^+(s)$ into the schema of a summarized vertex~$s$.

\begin{figure}
    \centering
    \includegraphics[width=0.5\textwidth,trim=4cm 10cm 8cm 3cm, clip=true]{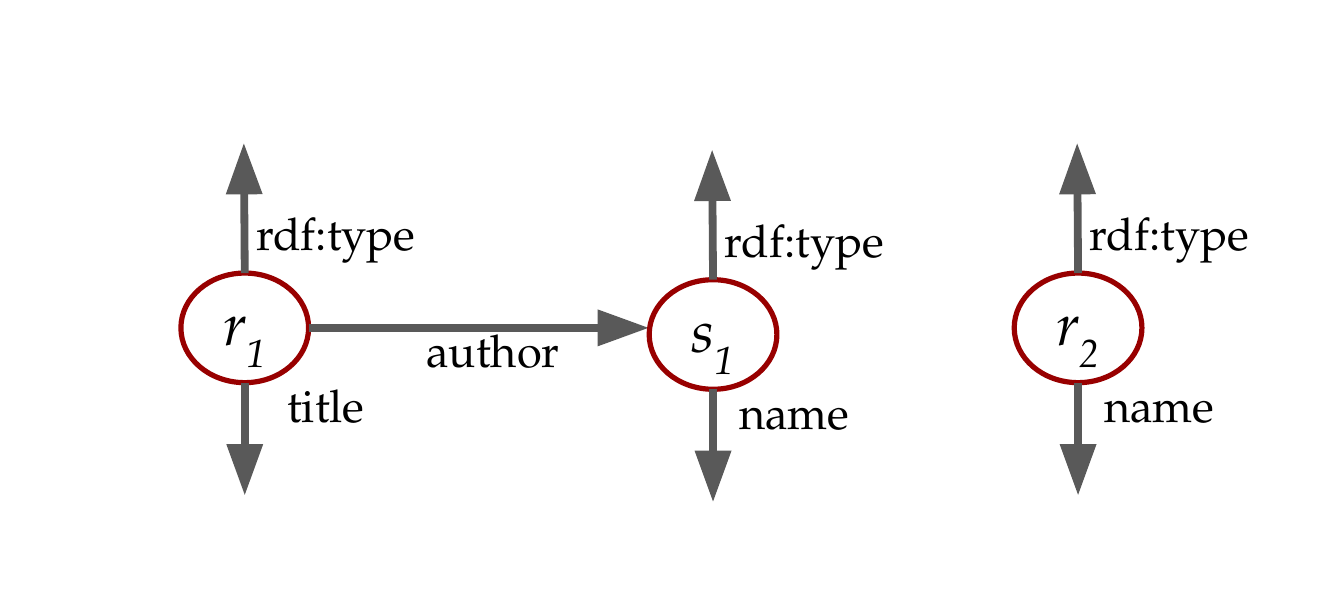}
    \caption{\label{fig:example-cse-2} Following the equivalence relation $CSE\mhyphen2$ defined in \cref{ex:cse2}, the two vertex summaries identified by the primary vertices $r_1$ and~$r_2$ summarize the vertices $v_1$ and~$v_2$ from \cref{fig:example-data-graph}, respectively. To represent the complex schema structure of $v_1$ in this vertex summary, we use a secondary vertex $s_1$ in addition to the primary vertex $r_1$ (see \cref{sec:graph-summarization}).}
\end{figure}

\begin{example}
\label{ex:cse-sse}
Let $CSE \mhyphen 1 = (\taut, \id, \id)$, where $\taut$~denotes the tautological equivalence relation, in which all vertices are equivalent and $\id$~denotes the identity equivalence relation.
$CSE \mhyphen 1$ considers vertices $s$ and~$s'$ equivalent iff, for every triple $(s, p, o)\in G$, there is a triple $(s', p', o')\in G$ such that $p'=p$ and $o'=o$. 
In other words, it is identical to the simple schema element $\POC$.
Similarly to the previous example, $\PC$ is identical to the CSE $(\taut, \id, \taut)$ and $\OC$ is the CSE $(\taut, \taut, \id)$. Formally, we may regard the simple schema elements as abbreviations for the corresponding CSEs; however, we implement them separately for efficiency.
\end{example}

\begin{example}
\label{ex:cse2}
Let $CSE \mhyphen 2 = (\taut, \id, \PC)$. 
We can see that this is a relaxation of $CSE\mhyphen1$ from \cref{ex:cse-sse}, in that $CSE\mhyphen1$ requires objects to be identical, whereas $CSE\mhyphen2$ only requires objects to have the same set of predicates.
In full, $CSE\mhyphen2$ declares vertices $s$ and~$s'$ to be equivalent iff for every triple $(s, p, o)\in G$, there is a triple $(s', p', o')\in G$ such that $p'=p$, and $o'$ and~$o$, in turn, have the same predicates, and vice versa.  
That is, for every predicate~$p$ that links $s$ to a neighbor~$o$, $p$~also links~$s'$ to a neighbor with the same set of predicates as~$o$, and vice versa.
This means that $CSE \mhyphen 2$ summarizes vertices using, in part, the schema of neighboring vertices $\Gamma^+(s)$.
In \cref{fig:example-cse-2}, we illustrate the vertex summaries for the example data graph visualized in \cref{fig:example-data-graph} according to the $CSE \mhyphen 2$ equivalence relation.
\end{example}

\subsection{Parameterizations of Schema Elements}
\label{sec:parameters}
Simple and complex schema elements are the basic building blocks of \model{}.
\model{} provides six parameterizations to specialize the schema elements' behavior.
For example, a parameterization can be applied that ignores a certain set of predicates or constructs predicate paths with a specific length.
In the following sections, we define six parameterizations, which are again motivated by the analysis of existing graph summaries in \cref{sec:related-work}.
The first two parameterizations address the triple features identified in \cref{sec:triple-features}, the next two address the subgraph features from \cref{sec:sub-graph-features}, and the final two address the semantic rule features from \cref{sec:rule-features}.

\subsubsection{Label Parameterization}
\label{sec:label-param}
The label parameterization can be used to restrict schema elements to consider only specific predicates.
This allows us to define, \eg the type cluster $\OCtype$, an object cluster that only compares objects connected over the predicate \texttt{rdf:type}.
Hence, $\OCtype$ compares vertices based on type sets. 

\begin{definition}
\label{def:labelParameterization}
The \textbf{label parameterization} is a function $lp(SSE, P_r)$, which takes as input a simple schema element $SSE$ and a set of predicates $P_r \subseteq P$ and returns a schema element $SE_{P_r}$.
The returned schema element $SE_{P_r}$ is defined analogously to $SE$ (Definitions \ref{def:poc}, \ref{def:pc}, and~\ref{def:oc}) but each existential and universal quantifier is restricted to tuples $(s,p,o)$ with $p\in P_r$.
In detail, $lp(SE, P_r)$ defines vertices $s$ and~$s'$ to be equivalent iff, for every triple $(s, p, o)\in G$ with $p\in P_r$, there is a triple $(s', p', o')\in G$ with $p'\in P_r$ such that:
\begin{itemize}
    \item $SE=\POC$ and $p=p'$ and $o=o'$; or
    \item $SE=\PC$ and $p=p'$; or
    \item $SE=\OC$ and $o=o'$.
\end{itemize}
\end{definition}

Using the predicate \texttt{rdf:type}, the label-parameterized object cluster $lp(\OC, \lbrace \text{\texttt{rdf:type}} \rbrace)$ summarizes vertices that have the same set of vertices connected with the predicate \texttt{rdf:type}, \ie vertices with the same type sets.
To ease notation, we refer to this label-parameterized object cluster as the \textbf{type cluster} $\OCtype$.
Two vertices $s$ and $s'$ are equivalent under $\OCtype$, iff $\ts{s} = \ts{s'}$.
Analogously, we define the label-parameterized predicate cluster $lp(\PC, P \setminus \lbrace \text{\texttt{rdf:type}} \rbrace)$, which summarizes vertices that have the same predicates explicitly excluding the RDF specific \texttt{rdf:type} predicate.
We denote this label-parameterized predicate cluster as property cluster $\PCrel$.
Two vertices $s$ and $s'$ are equivalent according to the $\PCrel$, iff $\ps{s} = \ps{s'}$.
\subsubsection{Set Parameterization}
\label{sec:set-param}
The set parameterization $sp$ is applied to simple schema elements.
In addition to the requirements of the SSE itself, the set parameterization also requires that all predicate and/or objects must be element of a given set~$S$.
In contrast to the label parameterization, which just ignores predicates not in~$S$, the set parameterization says that two vertices are automatically equivalent if they have any predicate not in~$S$.

\begin{definition}
\label{def:set-parameterization}
The \textbf{set parameterization} is a function $sp(SSE, S)$, which takes as input a simple schema element $SSE$ and a set of IRIs $S \subseteq V_I$ and returns a simple schema element $SSE|_{S}$.
$SSE|_S$ defines an equivalence relation $EQR$ such that $(s, s')\in EQR$ iff
\begin{itemize}
  \item if $SSE\in\{\PC, \POC\}$, 
  \begin{itemize}
    \item $S\neq\emptyset$, and $s$ and $s'$ both have at least one outgoing edge; or
    \item for all $p,p'$, where there are triples $(s,p,o) \in G$ and $(s',p',o') \in G$, it must follow that $p,p'\in S$ and $s,s'$ are equivalent under $SSE$; or
    \item there are triples $(s,p,o) \in G$ and $(s',p',o') \in G$, where $p,p'\not\in S$
  \end{itemize}
  
\item if $SSE\in\{\OC, \POC\}$,
  \begin{itemize}
    \item $S\neq\emptyset$, and $s$ and $s'$ both have at least one outgoing edge; or
    \item for all $o,o'$, where there are triples $(s,p,o) \in G$ and~$(s',p',o') \in G$, it must follow that $o,o' \in S$ and $s,s'$ are equivalent under $SSE$; or 
    \item there are triples $(s,p,o) \in G$ and $(s',p',o') \in G$, where $o,o'\not\in S$
  \end{itemize}

\end{itemize}
\end{definition}

The set parameterization can be combined, \eg with the label parameterization. 
We demonstrate this below.
\begin{example}
\label{ex:set-param}
Let us apply the set parameterization to the type cluster $\OCtype$ using the empty set $\emptyset$.
The set-parameterized type cluster $\OCtype|_\emptyset$ splits the data graph into at most two equivalence classes.
The first equivalence class (vertex summary) contains all vertices that have empty type sets (no type information provided in the data graph) and the second equivalence class (vertex summary) contains all vertices that have at least one type, if any such vertices exist.
\end{example}

\begin{example}
\label{ex:set-param-2}
Let us apply the set parameterization to the property cluster $\PCrel$ using the set $ S = \{p_1,p_2\}$.
The set-parameterized property cluster $\PCrel|_S$ compares vertices $v,v'$ based on the same property set only if the property set is a subset of $S$.
Any vertex $v$ where there is a $p \in \ps{v}$ that is not in $S$ will be summarized by the same vertex summary. 
Also, the corner case $\ps{v} = \emptyset$ will be summarized by that same vertex summary.
The remaining vertices are summarized based on the equivalence of their property sets.
Thus, there can be no more than four vertex summaries for any input graph, \ie the ones corresponding to property sets $\{p_1\}$, $\{p_2\}$, and $\{p_1,p_2\}$, and the one that summarizes all other vertices. 
\end{example}

The set parameterization can be used like a filter.
When such a set $S$ of predicates and/or objects is provided, the graph summary may contain a vertex summary that summarizes all vertices that have predicates and/or objects not in~$S$.
More importantly, it contains vertex summaries with exactly these predicates and/or objects. 
Subsequently, one could either filter out these vertex summaries or all other vertex summaries.
In \cref{ex:set-param}, we may only want to visualize all vertices with a non-empty type set. 
In \cref{ex:set-param-2}, we may only want to visualize all vertices that exclusively use properties $p_1$ and/or~$p_2$.

\subsubsection{Chaining Parameterization}
\label{sec:chaining-param}
Complex schema elements take the schema of two directly connected vertices into account.
The chaining of $k$~complex schema elements extends this to vertices within distance~$k$.
As discussed in \cref{sec:related-work}, this corresponds to a stratified $k$-bisimulation.
The chained complex schema element is denoted by $CSE^k$.

\begin{definition}
\label{def:chaining-parameter}
The \textbf{chaining parameterization} is a function $cp(CSE, k)$, which takes a complex schema element $CSE:=(\sim^s, \sim^p, \sim^o)$ and a chaining parameter $k \in \mathbb{N}_{>0}$ as input and returns an equivalence relation $CSE^k$ that corresponds to recursively applying $CSE$ to a distance of $k$~hops.  
$CSE^k$ is defined inductively as follows:
\begin{align*}
    CSE^1 &= (\sim^s, \sim^p, \sim^o)\\
    CSE^{k+1} &= (\sim^s, \sim^p, CSE^k)\qquad\text{for }k\geq 1\,.
\end{align*}
\end{definition}

\begin{example}
\label{ex:chaining}
Recall $CSE \mhyphen 2 := (\taut, \id, \PC)$ from \cref{ex:cse2} and denote by $\sim_2$ the equivalence relation it defines. 
Consider the chained complex schema element $CSE\mhyphen3 := cp(CSE\mhyphen2, 2)$. 
By \cref{def:chaining-parameter}, this corresponds to the CSE $(\taut,\id,\sim_2)$. 
Let $CSE\mhyphen3$ define the equivalence relation~$\sim_3$ and consider vertices $s$ and~$s'$.
We have $s\sim_3 s'$ iff the following conditions are met.
\begin{itemize}
    \item $s$ and $s'$ must have the same predicates, because predicate equivalence is defined by equality using the identity equivalence $\id$.
    \item For every neighbor $o \in \Gamma^+(s)$ via any predicate, there is a neighbor $o' \in \Gamma^+(s')$ via the same predicate, such that the vertices $o$ and $o'$ are equivalent under~$CSE\mhyphen2$. That is, for each predicate~$p$ that links $o$ to a neighbor with some predicates, the predicate~$p$ also links $o'$ to a neighbor with the same predicates.
    \item Vice versa, for every neighbor $o'$ of $s'$, a similarly corresponding neighbor $o$ of $s$ exists.
\end{itemize}
Intuitively, $CSE\mhyphen3$ determines the equivalence of vertices based not only on their neighbors but also on their neighbors' neighbors. 
As the chaining parameter~$k$ increases, we consider wider neighborhoods.
\end{example}

\begin{figure}[b]
    \centering
    \includegraphics[width=0.6\textwidth,trim=4cm 11cm 4cm 3cm, clip=true]{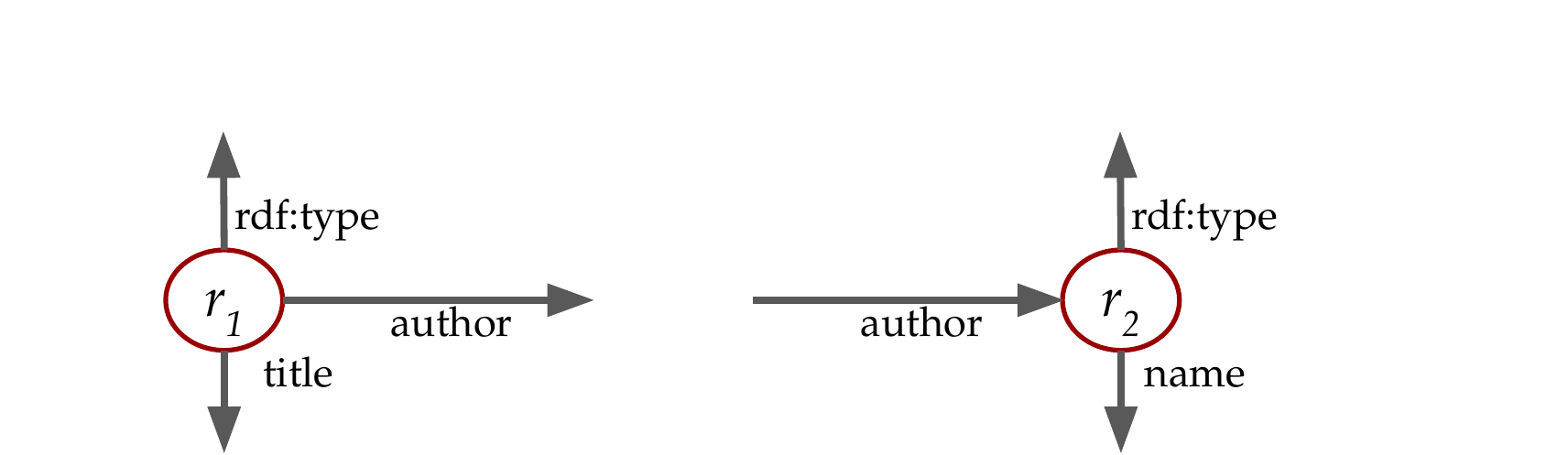}
    \caption{\label{fig:example-bi-predicate-cluster} Following the $\bidir\mhyphen\PC$ equivalence relation, the two vertex summaries identified by the primary vertices $r_1$ and~$r_2$ summarize the vertices $v_1$ and~$v_2$ from \cref{fig:example-data-graph}, respectively.}
\end{figure}

\subsubsection{Direction Parameterization}
\label{sec:direction-param}
The direction parameterization $dp$ is applied on schema elements to use only outgoing predicates (parameter $\delta = \outdir$), only incoming predicates ($\delta=\indir$), or both ($\delta=\bidir$).
Our schema elements $\OC$, $\PC$, $\POC$, and CSE only take outgoing predicates into account.
Schema structures like Characteristic Sets~\cite{DBLP:conf/icde/NeumannM11} use also incoming predicates.
To address incoming predicates, a parameterized version of the three simple schema elements can be defined by using the incoming triples $(x, p, v) \in G$ with the summarized vertex $v$ in object position.
\begin{definition}
\label{def:direction-param}
The \textbf{direction parameterization} is a function $dp(SE, \delta)$, which takes as input a schema element $SE$ and one direction parameter $\delta \in \{\indir, \outdir, \bidir\}$ and returns a schema element $\delta\mhyphen SE$, respectively.
Direction~$\indir$ uses only incoming predicates, $\outdir$~uses only outgoing predicates, and $\bidir$ uses both, incoming and outgoing predicates.
The returned schema element $\delta\mhyphen SE$ is a modification of $SE$ in which all assertions made by $SE$ about the triples are applied on the incoming edges only, the outgoing edges only, or on both.
This means that $\outdir\mhyphen SE = SE$, $\indir\mhyphen SE$ is $SE$ modified to use incoming predicates instead of outgoing and $\bidir\mhyphen SE = \indir\mhyphen SE \cap \outdir\mhyphen SE$.
\end{definition}
\begin{example}
The bidirectional predicate cluster $\bidir\mhyphen\PC$ partitions the data graph by comparing the triples based on common incoming and outgoing predicates.
Thus, the equivalence relation $\bidir\mhyphen\PC$ holds true, iff for each triple $(s_1,p_1,o_1) \in G$ there exists a triple $(s_2,p_1,o_2) \in G$, and vice versa, and for each triple $(s_3,p_3,s_1) \in G$ there exists a triple $(s_4,p_3,s_2) \in G$, and vice versa.
The bidirectional predicate cluster $\bidir\mhyphen\PC$ is visualized in \cref{fig:example-bi-predicate-cluster}.
\end{example}

\subsubsection{Inference Parameterization}
\label{sec:rdfs-param}
\begin{figure}[t]
\centering
    \includegraphics[scale=0.55,trim={3cm 9cm 4cm 5cm}]{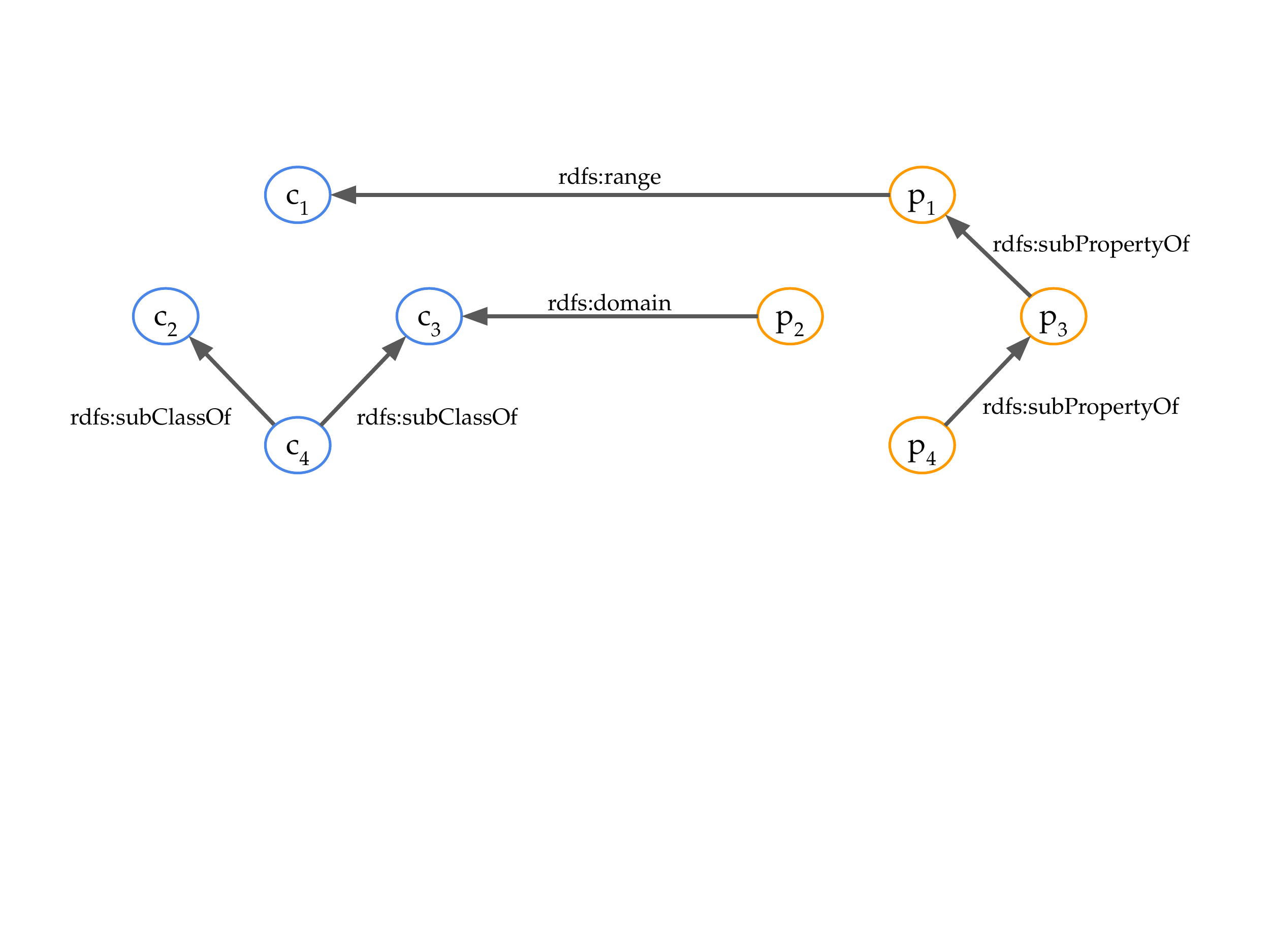}
    \caption{\label{fig:vocabulary-graph}Simple example of a RDFS vocabulary graph. Each vertex represents either a type or a property. To infer implicit information, starting from a given type or property vertex, inferrable information is collected by following edges. The rules on how to collect inferrable information are defined in the RDFS standard~\cite{w3c-rdf-schema}. }
\end{figure}

Some graphs contain semantic information, \eg ontologies described with the RDF Schema (RDFS) vocabulary which, like RDF, is standardized by the W3C~\cite{w3c-rdf-schema}.
The inference parameterization is the first parameterization to include the semantics of the data graph in the structural graph summary and also falls under our category of semantic rule features (see \cref{sec:rule-features}).
The inference parameterization $op$ is applied on the graph~$G$ and enables ontology reasoning using a vocabulary graph~$VG$.
Applying RDFS inferencing means that a vocabulary graph $\VGRDFS$ is constructed from the data graph~$G$, which stores all hierarchical dependencies between types and properties denoted by RDFS properties found in the data graph~$G$.
To construct the RDFS vocabulary graph, we first extract all triples containing RDFS vocabulary terms, namely all properties
$\PRDFS = \lbrace\texttt{rdfs:subClassOf}, \texttt{rdfs:subPropertyOf}, \texttt{rdfs:range}, \texttt{rdfs:domain}\rbrace$.
Subsequently, we add the corresponding hierarchical dependencies of \texttt{rdfs:subClassOf} and \texttt{rdfs:subPropertyOf} in a polytree\footnote{A polytree is an orientation of an undirected tree.} structure with further cross connections regarding \texttt{rdfs:range} and \texttt{rdfs:domain}~\cite{DBLP:conf/gvd/2018}.
An example of such a vocabulary graph is illustrated in \cref{fig:vocabulary-graph}.

\begin{definition}
An \textbf{RDFS vocabulary graph} is an edge-labeled directed multigraph $\VGRDFS \subseteq (V_C \cup P) \times \PRDFS \times (V_C \cup P)$.
The set of vertices is the union of types $V_C$ and predicates $P$ in $G$.
The edges are labeled with predicates $p \in \PRDFS$.
\end{definition}

%

With hierarchical dependencies of types and predicates represented using our vocabulary graph, additional triples can be inferred using our inference parameterization.
For a given triple in the data graph~$G$, we can look up the vertex representing the used predicates or type in $VG$. 
Starting from this vertex in $VG$, the information needed for inference is contained in a polytree with the vertex itself as root.
The entailment rules for $\VGRDFS$ are defined by the W3C~\cite{w3c-rdf-schema}. 
A comprehensive overview of these rules is presented in~\cite{DBLP:books/daglib/0028543}.

%

%
\begin{definition}
The \textbf{inference parameterization} is a function $op(G, VG)$ which takes any data graph~$G$ and a vocabulary graph~$VG$ as input and, based on the entailment rules defined in~$VG$, returns a data graph $G_{VG}$, which also includes all inferred triples.
\label{def:inference-param}
\end{definition}

With our notion of inference, we can infer implicitly stated triples, \eg by adding the implicitly stated types and properties of vertices to the type sets and property sets of the respective vertices. 
For example, if a vertex has the type \texttt{Proceedings} and the vocabulary graph contains the triple (\texttt{Proceedings}, \texttt{rdfs:subClassOf}, \texttt{Book}), then the type \texttt{Book} can be inferred and added to the vertex' type set. 

\model's formalization does not pose any restrictions on when the actual inference happens.
In general, we can distinguish inference \emph{inside} the structural graph summary and \emph{outside} the structural graph summary.
Inference inside means that all inferrable information is added to the original graph and the graph summary is computed for the new data graph.
However, the order of these two is interchangeably. 
Inferring on the data graph and then summarizing can be equivalent to summarizing the data graph and then inferring on the graph summary~\cite{DBLP:conf/semweb/LiebigVOM15,DBLP:journals/vldb/GoasdoueGM20}.
The latter one then may require another update on the graph summary. 

Inference outside means that we compute the graph summary for the original graph and keep the constructed vocabulary graph.
When we query the graph summary to fulfill our task, we can infer only the additional information that affects the query result, \eg by generating additional queries based on entailment rules defined in the vocabulary graph.
The result sets for inferencing inside and outside are equivalent.
This decision affects the build-time and the size of the computed structural graph summary as well as time to compute query results on the structural graph summary.
We discuss practical implications of this decision in the empirical analysis in \cref{sec:dataset-analysis}.

\subsubsection{Instance Parameterization}
\label{sec:instance-param}
In RDF graphs, vertices and all of their outgoing triples are commonly referred to as RDF instances. 
Referencing this terminology, we define the instance parameterization $ip$.
This allows us to define rules to join sets of outgoing triples of different vertices, \eg vertices linked with \texttt{owl:sameAs}.
The instance parameterization is the final parameterization regarding the rule features (see \cref{sec:rule-features}).
\begin{definition}
The \textbf{instance parameterization} $ip$ is a function $ip(SE, \Delta)$, which extends any schema element $SE$ to additionally take the schema of all equivalent vertices into account, following the instance equivalence relation $\Delta$.
The returned schema element $SE[\Delta]$ is an extension of $SE$, which restricts the triples to be in $[v]_{\Delta}$.
Thus, $ip$ merges vertices that are equivalent under~$\Delta$ and then applies $SE$.
\label{def:instance-param}
\end{definition}

In the following, we give two definitions of such an instance equivalence relation parameter $\Delta$, namely SameAs instances $\sigma$ and related property instances $\rho$.

\paragraph{SameAs instances $\sigma$}
In the context of Linked (Open) Data, the \texttt{owl:sameAs} property is of particular interest since $(s,\text{\texttt{owl:sameAs}},s')$ explicitly states the equivalence of the entities identified by the vertices $s$ and $s'$.
To take this information into account, we use the notion of \textbf{SameAs instances} $[v]_{\sigma}$, which are equivalence classes of vertices or, equivalently, unions of vertices.
Two vertices ${s}$ and~${s'}$ are equivalent according to the equivalence relation $\sigma$, iff there is a property-path in the data graph~$G$ labeled \texttt{owl:sameAs} from $s$ to $s'$~\cite{DBLP:conf/gvd/2018}.
Note that the property-path is independent of the direction of the \texttt{owl:sameAs} relation.
When summarizing vertices using the SameAs instance parameterization, we take the schema information from all equivalent vertices into account.

\begin{figure}[!t]
\centering
    \includegraphics[scale=0.55,trim={7cm 12cm 4cm 5cm}]{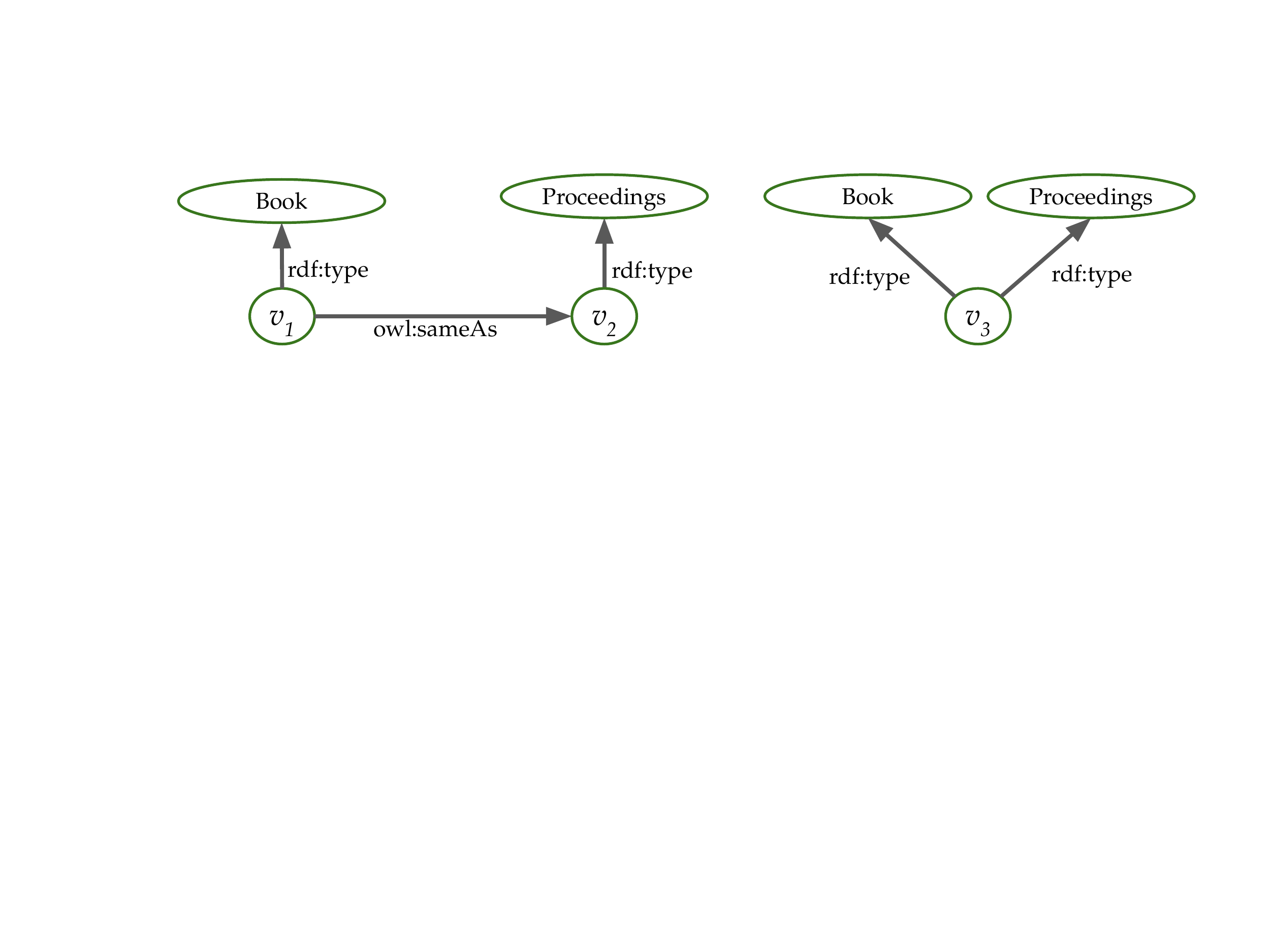}
    \caption{Sample data graph containing three vertices $v_1$, $v_2$, and~$v_3$.
The object cluster assigns $v_1$, $v_2$, and~$v_3$ to different vertex summaries.
The instance parameterized object cluster using SameAs instances $[v]_\sigma$ summarizes $v_1$, $v_2$, and~$v_3$ into a single vertex summary.}
\label{fig:sameAsInstanceSet}
\end{figure}

\begin{example} See \cref{fig:sameAsInstanceSet}.
According to the object cluster definition, the vertices $v_1$, $v_2$, and~$v_3$ are not equivalent since $v_1$ has the object \enquote{Book}, $v_2$ has the object \enquote{Proceedings}, and $v_3$ has the objects \enquote{Book} and \enquote{Proceedings}.
Merging $v_1$ and~$v_2$ to a SameAs instance $[v]_\sigma$ leads to the equivalence of all three vertices $v_1$, $v_2$, and~$v_3$.
\end{example}

\paragraph{Related property instances $\rho$}
Merging vertices also allows to model property cliques~\cite{DBLP:journals/vldb/GoasdoueGM20}.
Property cliques transitively check the co-occurrence of properties and summarize all vertices that have at least one property in common.
We can define an instance equivalence relation $\rho$ that leads to equivalent vertices if they share at least one property.
We call the equivalence classes $[v]_{\rho}$ \textbf{related property instances}.
We distinguish source related properties and target related properties~\cite{DBLP:journals/vldb/GoasdoueGM20}. 
Schema Elements parameterized with related property instances that consider outgoing properties use source related properties. 
For Schema Elements using outgoing properties, two vertices $s$ and~$s'$ are equivalent according to the equivalence relation $\rho$, iff there exists vertex $s'' \in [s]_{\rho}$ such that there are triples $(s'',p,o) \in G$ and $(s', p, o) \in G$ with $p \not= \texttt{rdf:type}$, or, similarly for some $s'''\in[s']_\rho$.

Schema Elements parameterized with related property instances that consider incoming properties use target-related properties. 
For Schema Elements using incoming properties, two vertices $s$ and~$s'$ are equivalent according to the equivalence relation $\rho$, iff there exists a vertex $s'' \in [s]_{\rho}$ such that there are triples $(x,p,s'') \in G$ and $(x', p, s') \in G$ and $p \not= \texttt{rdf:type}$, or similarly for some $s'''\in[s']_\rho$.

Note that $\rho$ takes transitively co-occurring properties into account.
Thus, $\rho$ may summarize two vertices that do not share any property in the data graph.

\subsection{Payload Elements}
\label{sec:payload-formal}
In total, \model{} provides four schema elements and six parameterizations.
They are designed to capture all schema structures defined by (semantic) structural graph summaries found in the related work and beyond.
In this section, we define several possible payload elements that are designed for purposes found in existing (semantic) structural graph summaries.

\begin{figure}[!t]
        \centering
    \includegraphics[width=.5\linewidth,trim={0cm 13cm 11cm 4cm}]{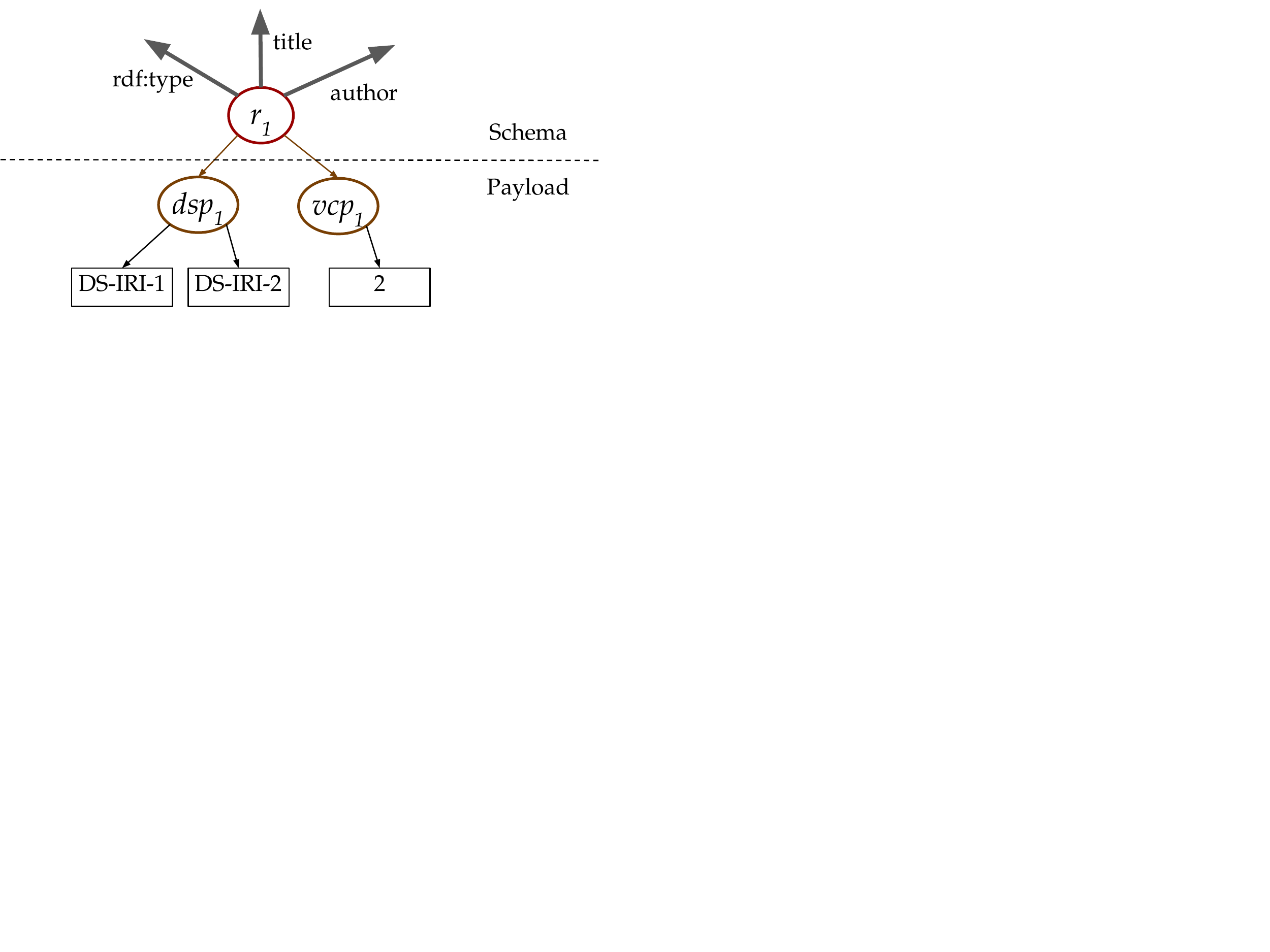}
    \caption{The vertex summary identified by the primary vertex $r_1$ summarizes vertices with the property set $\{\texttt{author}, \texttt{title},  \texttt{rdf:type}\}$. The data source payload $\datasourcepay_1$ contains information about the data sources and the vertex count payload element $\countpay_1$ contains the number of summarized vertices.}
\label{fig:payload-example}
\end{figure}

When computing the graph summary $SG$, information about the summarized vertices can be attached to vertex summaries by using the notion of payload~\citep{DBLP:conf/kcap/GottronSKP13}.
The payload contains information about the actual data, \eg number of vertices summarized or a reference to their data source.
Our intention is to make payloads as flexible as possible so we do not make any restrictions on what can be attached as payload and we allow multiple different kinds of payload to be attached to a single vertex summary.
To this end, we define $\PAY$ to be a set of payload elements. 
Payload elements map vertex summaries to a payload, \ie information extracted from the summarized vertices.
They are functions that define what information is attached.
For example, one payload element may attach the data sources of summarized vertices to vertex summaries while another payload element attaches the number of summarized vertices to vertex summaries. 
An example is illustrated in \cref{fig:payload-example}.
The payload $\datasourcepay_1$ contains data source IRIs that were extracted from the vertices summarized by the vertex summary represented by $r_1$. 
The payload $\countpay_1$ contains an integer value representing the number of vertices summarized by $r_1$.


\subsubsection{Indexing Vertices}
One purpose of structural graph summaries is to index vertices by their schema.
Thus, we need to be able to store the identifiers of summarized vertices as payload of schema elements.
%

\begin{definition}
Consider a graph summary of some graph, generated from an equivalence relation~$\EQR$.
The \textbf{vertex identity payload} is the set of identifiers (IRIs) of the vertices summarized by each vertex summary $vs$.
The vertex identity payload element $\vertexpay$ is a function that takes a vertex summary $vs_v$ of a vertex~$v$ as input and returns the set $[v]_\EQR$ of summarized vertices, \ie
$\vertexpay(vs_v) := [v]_\EQR$.
\label{def:instance-payload}
\end{definition}

\subsubsection{Query Size Estimation}
Another purpose of structural graph summaries is cardinality estimation for database queries~\cite{DBLP:conf/icde/NeumannM11}.
To implement this task, we only need the number of summarized vertices and not the vertex identifiers.
Thus, we can define another payload element, which maps only the integer value denoting the number of summarized vertices.

\begin{definition}
\label{def:instance-count}
The \textbf{vertex count payload} provides the number of vertices summarized by the vertex summary.
It is computed by the vertex count payload element $\countpay$, which takes a vertex summary $vs_v$ as input and returns an integer denoting the number of summarized vertices.
Thus, formally $\countpay$ is a function defined as $\countpay(vs_v) := |\vertexpay(vs_v)|$.
\end{definition}

\subsubsection{Data Source Selection}
\label{sec:payload}
Our third payload element is the data source payload element $\datasourcepay$, which is needed to implement the data search task~\cite{DBLP:conf/kcap/GottronSKP13}.
The data source payload element $\datasourcepay$ maps a vertex summary to the set of data source IRIs of all summarized vertices.
This payload is only valid for named graphs, \ie where each triple $(s,p,o)$ is extended to a tuple $((s,p,o), d)$ (see \cref{sec:graph-model}).

\begin{definition}
\label{def:datasource}
The \textbf{data source payload} contains the locations of the vertices summarized by the vertex summary.
The payload element $\datasourcepay$ is a function that takes a vertex summary $vs_v$ as input and returns all data source IRIs~$d$ for which there is tuple $((s,p,o), d)$ in~$G$ for some vertex~$s$ that is summarized by the vertex summary~$vs_v$.
Formally, the data source payload is defined as: 
\[\datasourcepay(vs_v) := \{d \mid \text{there is a tuple } ((s,p,o),d)\in G  \text{ for some }s\in \vertexpay(vs_v)\}\,.\]
\end{definition}

\begin{table*}[!b]
\caption{All graph summary models from the related work (see \cref{tab:index-model-features}) defined using the \model{} language.}
\label{tab:configurations}
{\renewcommand\arraystretch{1.3}
\setlength{\tabcolsep}{8pt}
\small
\begin{tabularx}{\textwidth}{p{0.cm} p{5cm} p{10.25cm}}
\toprule
& \textbf{Graph Summary Model}& \textbf{\model{} definition}\\
\midrule
\parbox[t]{4mm}{\multirow{4}{*}{\rotatebox[origin=c]{90}{\textit{Simple}}}}
& Attribute-based Collection~\cite{DBLP:conf/dexaw/CampinasPCDT12} & $(G, \PCrel, \{\vertexpay\})$\\
& Class-based Collection~\cite{DBLP:conf/dexaw/CampinasPCDT12} & $(G, \OCtype, \{\vertexpay\})$\\

& Characteristic Sets~\cite{DBLP:conf/icde/NeumannM11} & $(G, \mathrm{u}\mhyphen\PC, \{\countpay\})$\\

& SemSets~\cite{DBLP:conf/www/CiglanNH12} & $(G, \POC, \{\vertexpay\})$\\

\arrayrulecolor{black!50}\midrule

\parbox[t]{4mm}{\multirow{15}{*}{\rotatebox[origin=c]{90}{\textit{Complex}}}}
& LODex~\cite{lodex2015} & $(G, (\OCtype, \idrel, \OCtype), \{\vertexpay\})$\\

& Loupe~\cite{loupe2015} & $(G, (\OCtype, \idrel, \OCtype), \{\vertexpay\})$\\

& SchemEX~\cite{DBLP:journals/ws/KonrathGSS12} & $(G, (\OCtype, \idrel, \OCtype), \{\datasourcepay\})$ \\

& SchemEX+U+I~\cite{DBLP:conf/dexa/BlumeS20} & $(G_{\VGRDFS}, (\OCtype, \idrel, \OCtype)[\sigma], \{\datasourcepay\})$ \\

& ABSTAT~\cite{DBLP:conf/esws/SpahiuPPRM16a} & $(G_{\VGRDFS}, (\OCtype, \idrel, \OCtype), \{\vertexpay\})$\\

& TermPicker~\cite{DBLP:conf/esws/SchaibleGS16} & $(G, (\OCtype \cap \PCrel, \taut, \OCtype), \{\vertexpay\})$\\

\arrayrulecolor{black!20}\cline{2-3}

& Weak Summary~\cite{DBLP:journals/vldb/GoasdoueGM20} & $(G_{\VGRDFS}, \indir\mhyphen\PCrel[\rho] \OR \outdir\mhyphen\PCrel[\rho], \{\vertexpay\})$\\

& Strong Summary~\cite{DBLP:journals/vldb/GoasdoueGM20} & $(G_{\VGRDFS},\bidir\mhyphen\PCrel[\rho], \{\vertexpay\})$\\

& Typed Weak Summary~\cite{DBLP:journals/vldb/GoasdoueGM20}& $(G_{\VGRDFS}, (\OCtype|_\emptyset \cap ( \indir\mhyphen\PCrel[\rho] \OR \PCrel[\rho])) \OR \OCtype|_{V_C}, \{\vertexpay\})$\\

& Typed Strong Summary~\cite{DBLP:journals/vldb/GoasdoueGM20}& $(G_{\VGRDFS}, (\OCtype|_\emptyset \cap \bidir\mhyphen\PCrel[\rho]) \OR \OCtype|_{V_C}, \{\vertexpay\})$\\

\arrayrulecolor{black!20}\cline{2-3}
& Tran \etal~\cite{DBLP:journals/tkde/TranLR13}& $(G, (\taut,\id_{L},\taut)^k, \{\vertexpay\} )$\\

& A($k$)-index~\cite{DBLP:conf/icde/KaushikSBG02}& $(G, (\mathrm{u}\mhyphen \PC,\taut,\taut)^k, \{\vertexpay\} )$\\
& T-index~\cite{DBLP:conf/icdt/MiloS99}& $(G, (\mathrm{i}\mhyphen \PC,\taut,\taut)^k, \{\vertexpay\} )$\\

& Consens \etal~\cite{DBLP:journals/pvldb/ConsensFKP15}&  $(G, (\OCtype,\idrel,\OCtype)^k, \{\vertexpay\} )$\\

& Schätzle \etal~\cite{DBLP:conf/sigmod/SchatzleNLP13}& $(G, (\OC,\id, \OC)^k, \{\vertexpay\} )$\\

\arrayrulecolor{black!100}\bottomrule
\end{tabularx}
}
\end{table*}

\subsection{Defining the Existing Graph Summary Models with \model{}}
\label{sec:modeling-with-fluid}
In this section, we review all existing works analyzed in \cref{sec:related-work} and show how to define them using \model{}.
Thus, we demonstrate that \model{} can indeed be used to model all the existing (semantic) structural graph summaries in \cref{tab:index-model-features}.
We briefly describe their definitions using \model{}, which are shown in
\cref{tab:configurations}.
Note that the summaries defined in the related work were defined for specific tasks (e.g., query size estimation) so, here, we use the payload appropriate to that task. We could, of course, adapt the summaries to other tasks by using other payloads.

The first group of \textit{simple} graph summaries are defined using simple schema elements and parameterizations of them.
For Attribute-based Collection, we use the property cluster $\PCrel$, \ie label-parameterized predicate cluster not considering the predicate \texttt{rdf:type}.
Thus, we summarize vertices solely based on common outgoing properties (the property sets).
For Class-based Collection, we use the type cluster $\OCtype$, \ie the label-parameterized object cluster considering only the predicate \texttt{rdf:type}.
Thus, we summarize vertices solely based on common type sets.
For both graph summaries, we attach identifiers of the summarized vertices as payload using the vertex identity payload element $\vertexpay$.

Characteristic Sets are defined using the bidirectional predicate cluster $\bidir\mhyphen\PC$.
This means that vertices are summarized based on having the same outgoing predicates and the same incoming predicates.
As payload, we attach the number of summarized vertices using the vertex count payload element $\countpay$ to implement the cardinality estimation task.
SemSets are defined using the predicate-object cluster $\POC$, \ie vertices are summarized based on having the same outgoing triples.
As payload, we again attach the summarized vertices to implement related-entity search.

For the second group of \textit{complex} graph summaries, we use the complex schema element to combine equivalence relations.
To define SchemEX, we combine the type cluster $\OCtype$ and $\idrel$, the label-parameterized identity equivalence using all predicates except \texttt{rdf:type} (\ie only properties), in a complex schema element.
Since complex schema elements support predicate paths by default, they capture which predicate linked to which type set of which neighbor.
As payload, we use the data sources of summarized vertices, \ie the data source payload element $\datasourcepay$.
For SchemEX+U+I, we additionally include SameAs instances $\Delta = \sigma$ and an RDFS vocabulary graph $\VGRDFS$ on top of SchemEX. 
ABSTAT, LODex, and Loupe define the same graph summary as SchemEX, except that they use the vertex identity payload instead of the data source payload and ABSTAT uses RDF Schema inference on the data graph.
To define TermPicker, we use the intersection of the label-parameterized object cluster $\OCtype$ and the label-parameterized property cluster $\PCrel$.
The existence of any predicate equivalence ${\sim^p} \not= \taut$ in a complex schema element enables predicate paths, capturing which predicate links to which type set of a neighbor.
TermPicker defines a graph summary that does not use predicate paths, so we use the tautology equivalence~$\taut$ as predicate equivalence.
This way, no predicate paths are taken into account and all types of all neighbors are aggregated.

The next four summary models are proposed by Goasdou{\'{e}} \etal~\cite{DBLP:journals/vldb/GoasdoueGM20}.
All of their summary models use the instance parameterization with related property instances $\Delta = \rho$.
The Weak Summary summarizes vertices based on incoming properties or outgoing properties or both. 
Thus, we define it with directed property clusters $\indir\mhyphen\PCrel$ and $\outdir\mhyphen\PCrel$ combined using the extended union operator $\OR$.
The Strong Summary summarizes vertices based on incoming properties and outgoing properties.
Thus, we define it with a bidirectional property cluster $\bidir\mhyphen\PCrel$.
The Typed Weak Summary summarizes vertices based on the Weak Summary only if the vertices have empty type sets.
Otherwise, they are summarized based on having the same type sets. 
We express this using the set parameterization applied on the type cluster $\OCtype$ with $S = \emptyset$.
$\OCtype|_\emptyset$ partitions the vertices into those with no types and those with types in the data graph. 
$\OCtype|_{V_C}$ partitions the vertices into those with no types and those with the same type sets in the data graph.
Analogously, we define the Typed Strong Summary, which summarizes based on the Strong Summary only if the vertices have empty type sets. 
%


The final five approaches use $k$-bisimulation and can be defined using the chaining parameterization.
For the label- and height-parameterized graph summary model proposed by Tran \etal~\cite{DBLP:journals/tkde/TranLR13}, we use the complex schema element that only compares outgoing predicates included in the label set~$L$ up to maximum hop length of~$k$.
The parameters $L$ and $k$ are user defined.
Kaushik \etal~\cite{DBLP:conf/icde/KaushikSBG02} define their A($k$)-index using forward and backward bisimulation, \ie it follows incoming and outgoing predicate paths up to a maximum hop length of $k$.
We define this using the bidirectional predicate cluster chained in a complex schema element.
Analogously, we define the T-index proposed by Milo and Sucio~\cite{DBLP:conf/icdt/MiloS99} for incoming predicates only.
Consens \etal~\cite{DBLP:journals/pvldb/ConsensFKP15} summarize vertices based on outgoing property paths only and also compare the type sets of each vertex.
The graph summary proposed by Schätzle \etal~\cite{DBLP:conf/sigmod/SchatzleNLP13} is analogous to the previous one, but compares the vertices' identities instead of their type sets.

\subsection{Summary}
In summary, we define graph summaries to be 3-tuples of consisting of a data graph $G$, an equivalence relation $\EQR$, and payload elements $\PAY$.
We introduced three simple and one complex schema elements as well as six parameterizations to define graph summaries.
We demonstrated that these elements and parameterizations can be flexibly combined to define existing (semantic) structural graph summaries.
All analyzed graph summaries discussed in \cref{sec:related-work} can be expressed using the \model{} language.
We can also adapt existing graph summaries to new tasks.
The simplest modification is changing the payload elements, which is easily done, but has a big impact on the size of graph summary and on which tasks can be fulfilled.
For example, attaching the number of summarized vertices requires less space than attaching identifiers of summarized vertices.
Furthermore, we can adapt the captured schema structure by filtering out specific types and properties or enabling inference on semantic graphs. 
For example, we can define a new graph summary $\text{PC-Inferenced} := (G_{\VGRDFS}, (\taut, \idrel, \taut), \{\datasourcepay\})$ that only uses outgoing properties and uses inference.

There are various new possibilities to combine \model 's schema elements and parameterizations to define existing and new (semantic) structural graph summaries.
It can be shown that every combination of \model 's schema elements and parameterizations actually defines a partition over the data graph and thus, is a valid graph summary.
All schema elements and parameterizations are defined as equivalence relations.
The intersection and the extended union of such equivalence relations is again an equivalence relation.

FLUID provides a handful of elements and parameterizations to define (semantic) structural graph summary models. This
allows us to define a generic algorithm to compute (semantic) structural graph summaries. We present this algorithm and
analyze its computational complexity in the next section.

\section{Graph Summarization Algorithm}
\label{sec:complexity}
In this section, we introduce our algorithm to compute structural graph summaries.
We propose a single, parameterized algorithm, which can compute all structural graph summaries defined with \model{}. 
To compute a structural graph summary, we need to summarize vertices to vertex summaries, partitioning the data graph into disjoint subsets of vertices.
This is a version of the \textit{set union problem}, for which 
Tarjan's algorithm has been proven to be asymptotically optimal~\cite{DBLP:journals/jacm/TarjanL84}.
This algorithm is based on operations ``make-set'' and ``find'' and takes time $\Theta(n\cdot \alpha(n,n))$ for a sequence of $n$ of these operations, where $\alpha$ is the inverse of Ackermann's function. Although unbounded, $\alpha$~grows extremely
slowly and it is generally accepted that $\alpha(n,n) \leq 4$ for all practically possible inputs~\cite{DBLP:journals/jacm/TarjanL84}. We therefore refer to the running time of $\Theta(n\cdot \alpha(n,n))$ as being ``essentially linear time''.

\subsection{Algorithm}
\label{sec:algorithm}
\model{} describes the rules of how to combine schema elements, parameterizations, and payload elements.
To compute structural graph summaries, we propose an algorithm based on hash maps.
We assume that we can implement make-set and find operations in constant time (amortized) using hash maps, using hashes that are long enough to avoid collisions~\cite{DBLP:journals/ws/KonrathGSS12}.
Our algorithm is presented in \cref{algo:sequential-algorithm}.
To simplify the code, we only show an excerpt of the complete algorithm. 
Furthermore, we define a set of globally accessible hash maps in the data structure \emph{GlobalVariables} to ease readability.

In the first pass over the data graph~$G$ (\cref{algo:iterate-graph-start} to \cref{algo:iterate-graph-end}), we iterate once over all triples to prepare our data structures. 
Each triple using a property $p \in \PRDFS$ as predicate is added to the vocabulary graph $\VGRDFS$.
The remaining triples are added to the $\mathrm{InMap}$ and the $\mathrm{OutMap}$. 
The $\mathrm{InMap}$ stores the triples with the subject~$s$ as key and the $\mathrm{OutMap}$ stores triples with the object~$o$ as key.
This allows us to quickly retrieve all incoming and outgoing triples for each vertex in~$G$.
In case we are using SameAs instances, we construct sets of equivalent vertices following the \texttt{owl:sameAs} property in \cref{algo:sameas-instances-1,algo:sameas-instances-2}.
In case we are using related property instances, we construct sets of equivalent vertices following any outgoing or incoming property in \cref{algo:src-related-instances,algo:trg-related-instances}, respectively.

In the second pass (\cref{algo:iterate-inferencing-start} to \cref{algo:iterate-inferencing-end}), we infer incoming and outgoing triples according to the constructed Vocabulary Graph $\VGRDFS$. 
The inferred incoming and outgoing triples are stored in the hash maps $\mathrm{InfInMap}$ and $\mathrm{InfOutMap}$, respectively.
In the third pass, the schema for each vertex is computed.
For each (parameterized) schema element used to define the equivalence relation $\EQR$, we extract the schema and compute the vertex summaries.
These vertex summaries are stored in $\mathrm{SchemaMap}$ and identified by their primary vertices (see \cref{sec:graph-summarization}). 
Using the primary vertices, the summaries will be combined to form the vertex summary $vs$ for $\EQR$ (compare \cref{ex:cse2}). 
The primary vertices of all combined vertex summaries will become secondary vertices of the resulting summary $vs$.
Finally, for each payload element in $\PAY$, we compute the payload for each vertex $v$.
The resulting vertex summary $vs$ for each $v$ along with the payloads are added to the summary graph~$SG$ in \cref{algo:payload}.

\begin{algorithm}[h]
\scriptsize
\SetAlgoLined
\SetFuncSty{textsc}
\SetKwProg{Fn}{function}{}{end}
\SetKwInput{Input}{Input}
\SetKw{Result}{returns}
\SetKw{return}{return}
\SetKwFunction{compute}{ComputeGraphSummary}
\SetKwFunction{addElement}{AddElement}
\SetKwFunction{extract}{ExtractSchema}
\SetKwFunction{merge}{Merge}
\SetKwFunction{get}{Get}
\SetKwFunction{put}{Put}
\SetKwFunction{payload}{ExtractPayload}
\SetKwFunction{infer}{InferOntologyInformation}
\SetKwFunction{keys}{Keys}

\SetKwProg{struct}{struct}{$\ = \lbrace$}{$\rbrace$}

\Fn{\compute{$G,\EQR, \PAY$}}{
\Result{graph summary $SG$}
\BlankLine
\struct{GlobalVariables}{
    OutMap $\gets \emptyset$\; 
    InMap $\gets \emptyset$\; 
    InfOutMap $\gets \emptyset$\; 
    InfInMap $\gets \emptyset$\; 
    SameAsInstanceMap $\gets \emptyset$\; 
    SrcRelatedMap $\gets \emptyset$\; 
    TrgRelatedMap $\gets \emptyset$\; 
    SchemaMap $\gets \emptyset$\; 
}
$\rbrace$\;
\ForAll{$(s,p,o) \in G$}{\label{algo:iterate-graph-start}
    \eIf{$p \in \PRDFS $}{
    \tcc{Add triple to vocabulary graph if it is an RDF Schema triple.}
        $VG$.\addElement{$(s,p,o)$}\;\label{algo:construct-vocabulary-graph}
    }{
        \tcc{Index triples by subject vertex $s$.}
        OutMap.\merge{$s, \lbrace (s,p,o) \rbrace$}\;
        \tcc{Index triples by object vertex $o$.}
        InMap.\merge{$o, \lbrace (s,p,o) \rbrace$}\;
        \tcc{In case SameAs instances are used}
        \If{$p = \texttt{owl:sameAs}$}{
            SameAsInstanceMap.\merge{$s, \lbrace o \rbrace)$}\; \label{algo:sameas-instances-1}
            SameAsInstanceMap.\merge{$o, \lbrace s \rbrace$}\; \label{algo:sameas-instances-2}
        }
        \tcc{In case property-related instances are used, check for source-related properties...}
        SrcRelatedMap.\merge{$p, \lbrace s \rbrace$}\;\label{algo:src-related-instances}
        \tcc{... and check for target-related properties}
        TrgRelatedMap.\merge{$p, \lbrace o \rbrace)$}\;\label{algo:trg-related-instances}
    }
}\label{algo:iterate-graph-end}
\If{$G$ uses RDF Schema inferencing}{\label{algo:iterate-inferencing-start}
    \ForAll{$(v, OUT) \in $ OutMap}{
        \tcc{Prepare for inferencing (see \cref{sec:rdfs-param})}
        InfOutMap.\put{$v, VG$.\infer{$OUT$}}\;
    }    
    \ForAll{$(v, IN) \in $ InMap}{
        \tcc{Prepare for inferencing}
        InfInMap.\put{$v, VG$.\infer{$OUT$}}\;
        
    }
}\label{algo:iterate-inferencing-end}

\tcc{Design decision to only consider vertices that are the subject of at least one triple}
\ForAll{$v \in $ OutMap.\keys{}}{\label{algo:iterate-instances}
    \tcc{Extract the schema according to $\EQR$}
    $vs \gets $ \extract{$v, \EQR$}\;
    
    \ForAll{$pay \in \PAY$}{
        $p_i \gets pay$.\payload{$v$}\;
    }
    $SG$.\addElement{$vs, \bigcup_i p_i$}\;\label{algo:payload}
}

\return{$SG$}\;
}
\caption{Sequential, parameterized algorithm to compute structural graph summaries defined as equivalence relation $\EQR$ with the \model{} language}
\label{algo:sequential-algorithm}
\end{algorithm}

\begin{example}
\label{ex:PC-with-hash}
Suppose we want to extract the schema according to a predicate cluster ($\PC$) for a vertex~$s$ and there are two triples $(s,p_1,o_1)$, and $(s,p_2,o_2)$ in $G$ with $s$ as subject.
We extract the predicate for each triple and construct the corresponding set, \ie $\{p_1,p_2\}$.
We compute the hash value of the predicate set and store it in the schema hash map $\mathrm{SchemaMap}$.
The payload, \eg the number of summarized vertices, is stored as value.
For another vertex~$s'$ with $(s',p_1,o_3), (s',p_2,o_4) \in G$, the same predicate set is extracted, so we compute the same hash value.
When we update the $\mathrm{SchemaMap}$, we update the payload.
In this example, we increase the vertex counter by one.
Consequently, the resulting vertex summary defined as $\PC$ summarizes both vertices and only stores the defined payload.
\end{example}

One can see that the computationally expensive task is the extraction of the schema of each vertex, \eg extracting the property set in \cref{ex:PC-with-hash}.
However, our algorithm benefits from the fact that every schema element in \model{} is defined as equivalence relation.
In particular, for one vertex $s$ there is exactly one schema structure according to any one (parameterized) schema element, \eg one predicate cluster ($\PC$) and one label-parameterized object cluster ($\OCtype$) etc.
We define complex graph summaries by combining simple schema elements with complex schema elements, where we can include neighboring vertices in the schema structure.
For example, SchemEX uses type sets of neighboring vertices to the compute the schema of the actual vertex.
Therefore, one may assume that depending on the in-degree of a vertex, we have to extract the schema more than once.
Instead, we use a hash map to store, for each vertex in the data graph, all computed vertex summaries identified by secondary vertices.
This avoids the expensive task of extracting the schema of vertices more than once.

\begin{lemma}
\label{ass:only-once}
For each vertex $v$ in the data graph $G$, we need to compute the schema according to each schema element in the summary definition only once.
\end{lemma}

\subsection{Complexity Analysis}
We analyze the complexity of computing  (semantic) structural graph summaries defined with \model{}.
In order to estimate the computational complexity, we conduct a space and time analysis of the computation process for graph summaries defined with \model{}.
For this analysis, we assume that hash maps have amortized constant-time read and write access and that hashes are long enough to avoid collisions when hashing, \eg vertices' property sets and type sets~\cite{DBLP:journals/ws/KonrathGSS12}. 
Goasdou\'e \etal~\cite{DBLP:journals/vldb/GoasdoueGM20} justify this assumption, via their hypothesis~($\star$), and provide an alternative data structure using Tarjan's algorithm instead of hash maps.
In a previous work, we found evidence that supports their hypothesis, \ie that the numbers of incoming and outgoing predicates for each vertex are, in practice, bounded~\cite{CIKM-incremental-summary}.


For the space complexity of the summary, the important factor is how well the summary model summarizes the input graph, \ie how many different vertex summaries are needed to summarize all vertices in~$G$.
The worst case is that no two vertices are summarized by the same vertex summary, \ie all vertex summaries in the structural graph summary summarize exactly one vertex.
In this worst case, each vertex summary is a new entry for our hash map.
Thus, the upper bound for the time complexity and the upper bound for the space complexity are identical.
In the following, we analyze in more detail the influence of \model 's schema elements and parameterizations on the computational complexity.

\subsubsection{Schema Elements}
\label{sec:complexity:SEs}
Given a \model{} definition of a graph summary, we denote by $\simple$ the number of simple schema elements and by $\complex$ the number of complex schema elements used to define the equivalence relation $\EQR$.
The parameterizations are applied to these simple and complex schema elements and pose further restrictions or relaxations.

The data graph $G$ contains $n$ triples.
Simple schema elements determine whether two vertices $s,s'\in G$ are equivalent by considering only triples that have $s$ or~$s'$ as subject. 
To compute $\PC$, $\OC$, or $\POC$, we need to compute, for each vertex $s\in G$, the set $\{p\mid (s,p,o)\in G \text{ for some }o\}$, $\{o\mid (s,p,o)\in G \text{ for some }p\}$ or $\{(p,o)\mid (s,p,o)\in G\}$, respectively. 
All of these sets can be computed in a single scan of the graph. 
A hash for each set can be computed by combining the hashes of its elements, \eg using XOR or techniques based on symmetric polynomials~\cite{OKeefe}. 
The equivalence relations corresponding to the SSEs can be computed by comparing these hashes.

The complex schema element $CSE = ({\sim^s},{\sim^p},{\sim^o})$ (\cref{def:complex-schema-element}) summarizes vertices based on the equivalence relations $\sim^s$, $\sim^p$, and~$\sim^o$ which are, in turn, defined by simple or complex schema elements. 
We first compute the vertex summaries for these equivalence relations.
For each vertex $v \in G$ and for each equivalence relation, we store the primary vertex identifier of the vertex summary that summarizes~$v$. 
We now compute the vertex summary for $CSE$ in essentially the same way as we compute $\POC$ but, instead of vertex identifiers, we use primary vertices of corresponding vertex summaries. That is, for each vertex~$s$, we construct the set $\{(r_p,r_o)\mid (s,p,o)\in G\}$, where $r_p$ and~$r_o$ denote the primary vertices of the vertex summaries for $p$ and~$o$ under the equivalences $\sim^p$ and~$\sim^o$, respectively. 
Again, we hash this set by combining the hashes of its elements. 
The equivalence relation for $CSE$ is now computed by comparing these hashes and checking for equivalence under~$\sim^s$. 
To check whether $s\sim^s s'$, we check whether their vertex summaries have the same primary vertex.

Thus, without parameterizations, a \model{} expression containing $\simple$ SSEs and $\complex$ CSEs can be evaluated in time and space  $\mathcal{O}((\simple + \complex) \cdot n)$, on an input graph with $n$~triples.

\subsubsection{Label Parameterization}
The label parameterization reduces the number of considered objects and/or predicates for each simple schema element by restricting to those in the set~$P_r$. Thus, we compute the parameterized schema element as described in \cref{sec:complexity:SEs} except that, for each triple $(s,p,o)\in G$, we must check whether $p\in P_r$ and/or $o\in P_r$. This requires $\Theta(n)$ membership checks, each of which takes time $\mathcal{O}(|P_r|)$. Note that $|P_r|$ depends only on the parameterized SSE being evaluated and not on the data graph. Thus, with respect to the input data graph, the parameterized SSE is evaluated in time $\mathcal{O}(n\cdot |P_r|)$, which is linear in~$n$. The space cost remains linear in~$n$ since applying the parameterization might not change the graph summary that is produced.

\subsubsection{Set Parameterization}
The set parameterization differs from the label parameterization only in what is done with subject vertices that have predicates and/or objects outside the parameter set~$S$. By the same argument as above, the running time is $\mathcal{O}(n\cdot |S|)$ (where, again, $|S|$ is independent of the input graph) and the space used is linear in~$n$.


\subsubsection{Chaining Parameterization}
Let $CSE=(\sim^s,\sim^p,\sim^o)$ be a complex schema element. By \cref{def:chaining-parameter}, the chaining parameterization $cp(CSE, k)$ denotes an equivalence relation defined by recursively applying $CSE$ $k$~times. This is simply $k$~nested CSEs,
$(\sim^s, \sim^p, (\sim^s, \sim^p, (\cdots (\sim^s, \sim^p, \sim^o)\cdots)))$,
which can be evaluated in time and space $\mathcal{O}(k\cdot n)$.



\subsubsection{Direction Parameterization}
With the direction parameterization, incoming and/or outgoing triples can be used (\cref{def:direction-param}).
Any incoming predicate of a vertex $v$ is the outgoing property of another vertex $v'$.
Thus, for bidirectional schema elements $\bidir\mhyphen\EQR$, we may have to consider each triple twice.
However, this is still linear w.r.t.\@ to the number~$n$ of triples in the data graph.

\subsubsection{Inference Parameterization}
For the inference parameterization, we have to consider the space and time required to build both the summary graph $SG$ and the vocabulary graph $VG$.
As defined in \cref{sec:rdfs-param}, the vocabulary graph is constructed by adding types and properties to it, if there exists a triple in $G$ using a predicate in $\PRDFS$ (\cref{algo:sequential-algorithm}, \cref{algo:construct-vocabulary-graph}).
In the input data graph $G$ of size $n$, we find $r$ schema triples that use a property in $\PRDFS$, with $r \leq n$.
Constructing the vocabulary graph is done analogously by updating hash maps. 
Thus, constructing the vocabulary graph changes neither the build-time nor the space complexity.

However, inferring information can change the overall complexity.
For our analysis, we distinguish the two cases of adding inferrable information \emph{inside} the structural graph summary and adding inferrable information \emph{outside} of the structural graph summary (see \cref{sec:rdfs-param}).

The first case of inferencing inside can have a big impact on the build-time.
The complexity depends on the number~$t$ of additional triples that can be inferred for each triple in $G$ from the $r$ triples in the vocabulary graph with $t \leq r$.
Thus, applying the inference parameterization to a \model{} definition using $\simple$~SSEs and $\complex$~CSEs and no other parameterizations defines a graph summary model that can be computed in time and space $\Theta((\simple + \complex) \cdot n \cdot t)$.

Note that $t$ may itself be a function of~$n$, in which case $t = \mathcal{O}(n)$.
This means that, when using the inference parameterization, we have in the worst case running time and space usage $\mathcal{O}(n^2)$ for the case of inference inside.

\begin{example}
Consider an RDF graph with the triple set
\begin{align*}
    \{ & (s_1, p_1, o_1), (s_2, p_1, o_2), \dots, (s_{n/2}, p_1, o_{n/2}), \\
       & (p_1, \texttt{rdfs:subPropertyOf}, p_2), \\
       & (p_2, \texttt{rdfs:subPropertyOf}, p_3), \\
       & \vdots \\ 
       & (p_{n/2-1}, \texttt{rdfs:subPropertyOf}, p_{n/2})  \}\,,
\end{align*}
where we assume $n$ to be even. Inference on this graph results in the graph containing the $n^2/4$ triples $\{(s_i, p_j, o_i) \mid 1\leq i,j\leq n/2\}$. This shows that inference can produce a quadratic blow-up in the number of tuples in a graph.
\end{example}

In the subsequent section, we analyze large real-world semantic graphs obtained from the Linked Open Data cloud.
Our analysis suggests that adding inferrable information inside the structural graph summary may be infeasible in a practical setting.
In this case, adding inferrable information should be done outside of the structural graph summary.

The vocabulary graph can be implemented using hash maps, which guarantees amortized constant time for lookup and addition operations.
Inference operations are linear in the number of inferrable types and properties.
For static vocabulary graphs, caching can be used to improve the run time.
Thus, we have the same time complexity as for the space complexity for the case of inference inside.
For the second case of inference outside the structural graph summary, space and build-time complexity remain unchanged.
%

\subsubsection{Instance Parameterization}
The instance parameterization aggregates vertices to unions of vertices, \eg SameAs instances or related property instances.
The instance parameterization does not increase the space complexity since no new triples are added to the data graph.
This contrasts with the inference parameterization, which potentially adds new triples.
The instance parameterization can decrease the number of different vertex summaries in the graph summary. 

As described in \cref{sec:algorithm}, two vertices can be aggregated using the instance parameterization, in principal, in amortized constant time using hash maps (\cref{algo:sequential-algorithm}, \cref{algo:sameas-instances-1,algo:sameas-instances-2,algo:src-related-instances,algo:trg-related-instances}).
To handle transitivity, we have to recursively access the hash map $\mathrm{SameAsInstanceMap}$ (for SameAs instances) or $\mathrm{SrcRelatedMap}$ and $\mathrm{TrgRelatedMap}$ (for related property instances).
Transitivity means here that, for each equivalent vertex~$v'$ of some vertex~$v$, we use the hash maps to look up the set of equivalent vertices of~$v'$.
The transitive closure only needs to be computed once.

In the worst case, every vertex is equivalent to each other. 
In this case, we have $\mathcal{O}(n)$ lookup operations for the first vertex~$v$.
Since all vertices are equivalent, the computation terminates.
For each vertex $v$ that is aggregated using the instance parameterization, one fewer vertex summary needs to be computed.
Thus, the time complexity also does not increase.

\subsection{Summary}
Graph summaries defined with \model{} can be computed in linear time and space with respect to the number of triples~$n$, unless the inference parameterization is used, in which case the running time and space usage may be quadratic in~$n$.
%
For the inference parameterization, we run our own analyses on large real-world datasets to justify the assumption of essential linear runtime.
This analysis on very large semantic graphs appears in the next section.

\section{Empirical Analysis of the Inference Parameterization}
\label{sec:dataset-analysis}
To estimate the impact of RDF Schema inference in a practical setting, we analyze four large semantic graphs obtained from the Linked Open Data cloud.
The graphs have different characteristics resulting from different crawling strategies.
In particular, we selected graphs with different sizes (number of triples~$n$), obtained in different years, and containing multiple different data providers, \eg DBpedia, Wikidata, and others.
Note that we did not perform any pre-processing except removing invalid (not parsable) triples.

\subsection{Datasets}
The first dataset is called \timbl{} and contains about $11$ million triples~\cite{DBLP:journals/ws/KonrathGSS12}. 
The crawl was conducted in 2011 with a breadth-first search starting from the single URI of Tim Berners-Lee's FOAF profile.
The second dataset is called \dyldo{} and contains about $127$~million triples~\cite{Kaefer2013}. 
The Dynamic Linked Data Observatory (DyLDO) provides regular snapshots from the Linked Open Data cloud.
We use their first snapshot crawled in May 2012 starting from about $95,000$ seed URIs.
This crawl was done with a breadth-first search but limited to a crawling depth of two~\cite{Kaefer2013}.
Although there are more recent DyLDO snapshots, they are decreasing in size.
Thus, we decided to take the first and largest one.
The third dataset is the Billion Triple Challenge 2019 dataset (\btc{}), which contains about $2$~billion triples~\cite{DBLP:conf/semweb/HerreraHK19}.
The \btc{} dataset was crawled breadth-first in January 2019 starting from $450$ seed URIs taken from the DyLDO dataset~\cite{DBLP:conf/semweb/HerreraHK19}.
To the best of our knowledge, the largest collections of Linked Open Data is the LOD Laundromat dataset\footnote{\url{http://lodlaundromat.org/}} (\laundromat{})~\cite{DBLP:journals/semweb/RietveldBHS17}.
It contains more than $38$~billion triples and combines various other data sources into one single dataset.


\subsection{Procedure}
We analyzed the four datasets with respect to RDF Schema inference.
In the top half of \cref{tab:dataset-analysis}, we present the size of the computed vocabulary graph $\VGRDFS$ for each dataset.
Furthermore, we distinguish between property vertices, \ie vertices representing a property $p$ used in the data graph $G$, and type vertices.
As described in \cref{sec:rdfs-param}, properties and types used in the data graph~$G$ are only added to the vocabulary graph $\VGRDFS$ if there is a triple about this property or type with a property $p_r \in \PRDFS$.
For each such property $p_r$, we also counted the number of (hierarchical) relationships expressed, namely for \texttt{rdfs:subPropertyOf}, \texttt{rdfs:subClassOf}, \texttt{rdfs:domain}, and \texttt{rdfs:range}.

In the lower half of the table, we present data on the impact of RDFS inference on each dataset.
First, we counted the number of properties $p$ and the number of types $t$ in the datasets and distinguished whether they are represented as vertex in the vocabulary graph $\VGRDFS$, \ie have inferrable information, or not.
Note that we counted how many triples use the property $p$ as predicate, \ie $(s,p,o)$, or label a vertex with the type $t$ using \texttt{rdf:type} as predicate, \ie $(s,\text{\texttt{rdf:type}}, t)$.
This way, we can use these numbers to estimate an upper bound for the size of the structural graph summary.
Finally, we counted the number of additional properties and types that can be added to the dataset based on the inference rules in $\VGRDFS$.
The increase factor reflects for properties, types, and both, how much additional information is added.
These factors denote the influence of the inference parameterization on the build-time and space complexity of structural graph summaries for the analyzed datasets.
Note that while the factor denotes the increased build-time, the storage space for the structural graph summary might be significantly lower.
As described in \cref{sec:graph-summarization}, structural graph summaries summarize the data graph, \ie common schema structures are only stored once in the summary graph.

\begin{table*}[!t]
\small
\caption{\label{tab:dataset-analysis}Dataset analysis with rounded number to the nearest thousand $T$, million $M$, or billion $B$.}
\begin{tabularx}{\linewidth}{p{5.5cm} r r r r}
\toprule
\textbf{} & \textbf{\timbl{}} & \textbf{\dyldo{}}& \textbf{\btc{}} & \textbf{\laundromat{}}\\
\midrule

\textbf{RDFS Vocabulary Graph $\VGRDFS$} \\


Property vertices & $12T$ & $31T$ & $26T$ & $330T$ \\

\texttt{rdfs:subPropertyOf} triples & $10T$ & $39T$ & $38T$ & $211T$\\

\arrayrulecolor{black!50}\midrule

Type vertices & $141T$ & $312T$ & $339T$ & $3.7M$ \\

\texttt{rdfs:subClassOf} triples  & $644T$ & $3M$ & $2M$ & $144M$ \\
\texttt{rdfs:domain} triples & $10T$ & $20T$ & $19T$ & $204T$ \\
\texttt{rdfs:range} triples & $10T$ & $19T$ & $16T$ & $196T$\\

\arrayrulecolor{black!50}\midrule
Total vertices $|V|$ & $153T$ & $343T$ & $365T$ & $4M$\\
Total triples $|E|$ & $673T\ (6\%)$ & $3M\ (2\%)$ & $2M\ ({<} 1\%)$ & $145M\ ({<} 1\%)$ \\


%

\arrayrulecolor{black!100}\midrule

\textbf{Dataset and Inference} \\

Properties $p \not\in \VGRDFS$ & $889T\ (10\%)$ & $75M\ (66\%)$ & $98M\ \ (5\%)$ & $12B\ (42\%)$\\
Properties $p \in \VGRDFS$ & $8M\ (90\%)$ & $39M\ (34\%)$ & $1.9B\ (95\%)$ & $16B\ (58\%)$  \\
\emph{Properties in dataset} & $9M$ & $114M$ & $2.1B$& $28B$\\ 

Added through inference & $7M$ & $33M$ & $703M$ & $50B$\\

\emph{Total number of properties} & $17M$ & $148M$ & $2.7B$ & $78B$\\

\arrayrulecolor{black!50}\midrule

Types $t \not\in \VGRDFS$ & $181T\ \ (9\%)$ & $2.6M\ (20\%) $ & $20M\ (22\%)$ & $938M\ (23\%)$ \\
Types $t \in \VGRDFS$ & $2M\ (91\%)$ & $10M\ (80\%)$ & $71M\ (78\%)$ & $3B\ (77\%)$ \\
\emph{Types in dataset} & $2M$ & $13M$ & $92M$ & $4B$\\ 

Added through inference & $17M$ & $120M$ & $2.4B$ & $646B$\\
\emph{Total number of types} & $19M$ & $134M$ & $2.5B$ & $650B$ \\ 

\arrayrulecolor{black!50}\midrule

Increase factor for properties & $1.9$ & $1.3$ & $1.3$ &  $2.7$ \\
Increase factor for types & $9.2$ & $10.5$ & $27.5$& $160.3$  \\

Increase factor (total) & $\bm{3.3}$ & $\bm{2.2}$ & $\bm{2.5}$ & $\bm{22.2}$ \\

\arrayrulecolor{black!100}\bottomrule
\end{tabularx}
\end{table*}

\subsection{Results}
From the results of our analysis (\cref{tab:dataset-analysis}), we can state that the size of the computed vocabulary graph $\VGRDFS$ is only a small fraction of the input data graph $G$.
On average, the vocabulary graph is only about $5\%$ of the data graph in terms of number of triples.
In contrast, adding the inferrable information increases the size of the data graph by a factor of between $2$ and~$20$.
Notably, the largest dataset (\laundromat{}) also has by far the largest increase factor.

Another main result is that over all datasets, \texttt{rdfs:subClassOf} hierarchies are considerably longer than \texttt{rdfs:subPropertyOf} hierarchies.
While on average $1.04$ (max $304$) additional properties can be inferred, on average $13.64$ (max $3641$) additional types can be inferred.
This is also reflected in the number of \texttt{rdfs:subClassOf} triples in the vocabulary graphs.
Over all four vocabulary graphs, \texttt{rdfs:subClassOf} triples make up $95\%$ to $99\%$ of all triples in the vocabulary graph.

Furthermore, we have found that in three out of four datasets, more properties have inferrable information, \ie appear in $\VGRDFS$, than properties that do not have inferrable information.
On average, $70\%$ of the properties used in the dataset have inferrable information.
For types, we found that over all four datasets, on average more than $80\%$ have inferrable information.

\subsection{Discussion}
Our results indicate a considerable impact on the size of the data graph when using RDF Schema inference.
For three out of four datasets, the amount of data increases by a factor of about two to three.
A notable exception is here the \laundromat{} dataset.
Despite having a comparably small vocabulary graph (less than $1\%$ of data graph size), more than 640~billion additional types can be added.
This means, after inference, about 160 times more types appear in the graph.
The\laundromat{} is the only dataset in our analysis that is an aggregation of user-uploaded datasets.
In contrast, the other datasets are crawled from the Web using different strategies.

Moreover, the vocabulary graphs $\VGRDFS$ comprise only a small fraction of the data graph, thus, allowing a compact representation of RDF Schema information.
In our analysis, we found that for larger datasets, the size decreases compared to the dataset size.
Therefore, it appears practical to add inferrable information outside of the structural graph summary for large semantic graphs, \ie store structural graph summary and vocabulary graph separately.
However, Goasdou\'e \etal~\cite{DBLP:journals/vldb/GoasdoueGM20} also propose an optimization technique to improve the performance of inference, which improved the time needed to perform inference in graph summaries by up to $94\%$.
As noted in \cref{sec:rdfs-param}, inferring on the data graph and then summarizing is equivalent to summarizing the data graph and then inferring on the graph summary~\cite{DBLP:conf/semweb/LiebigVOM15,DBLP:journals/vldb/GoasdoueGM20}.

One should also note that the structural graph summary is designed to be orders of magnitude smaller than the original graph.
It remains to be evaluated if the graph summaries grow by the same factor as the dataset, when RDF Schema inference is used.
In a previous work, we compared the size of one graph summary with and and without inference~\cite{DBLP:conf/dexa/BlumeS20}.
For the structural graph summary that extends SchemEX with RDF Schema and \texttt{owl:sameAs}, we have found a notably smaller increase in size.
For the \timbl{} dataset, the structural graph summaries increase on average by a factor of $1.2$ (instead of $3.3$) and for the \dyldo{} dataset, the structural graph summaries increase on average by factor of $1.5$ (instead of $2.2$).
Thus, the size of the structural graph summaries does not necessarily grow by the same factor as the data graph grows.
But, we assume that the increase factor of the data graph defines an upper bound of the increase factor of the semantic structural graph summary.

%

\section{Conclusion}
\label{sec:conclusion}
We have presented \model{}, a common model to define (semantic) structural graph summaries.
We defined \model 's building blocks, \ie four schema elements and six parameterizations, and demonstrated that they are sufficient to define existing (semantic) structural graph summaries and beyond.
Furthermore, we presented our \model{} language, which can be used to define, implement, and evaluate structural graph summaries.
Moreover, graph summaries defined with \model{} can be typically computed in time and space $\Theta(n)$ w.r.t.\@ $n$, the number of triples in the data graph.
Expressing (semantic) structural graph summaries with only a handful of elements and parameterizations in the \model{} language allows us to implement a single, parameterized algorithm to compute (semantic) structural graph summaries.
This algorithm takes as input a data graph and a graph summary model defined using the \model{} language and returns the computed structural graph summary as output.
Our implementation is available under an open-source license.\footnote{\url{https://github.com/t-blume/fluid-framework}}
This way, extensions of \model{} in the future are possible.
\model{} computes exact structural graph summaries.
In the future, we could integrate statistical approaches such as frequent pattern mining~\cite{DBLP:conf/esws/VolkerN11}.

\ifanon
\section*{Acknowledgments}
This research was co-financed by the EU H2020 project ANONYMOUS (\url{http://www.sectet-project}) under contract no.~123456.
\else
\section*{Acknowledgments}
This research was co-financed by the EU H2020 project MOVING (\url{http://www.moving-project.eu/}) under contract no.~693092.
\fi



\section*{References}
\bibliographystyle{abbrvnat}
\ifanon
\bibliography{bibliography-anon}
\else
\bibliography{parameterized-schema-paper-tcs}
\fi

\end{document}